\documentclass[preprintnumbers,article,amsmath,amssymb,floatfix,10pt,prd,onecolumn,
superscriptaddress,nofootinbib]{revtex4}
\usepackage{bbm}
\usepackage{amsfonts}
\usepackage{mathrsfs}
\usepackage{latexsym}
\usepackage{epsfig}
\usepackage{epstopdf}
\usepackage{graphicx}
\usepackage{amssymb}
\usepackage{amsmath}
\usepackage{dcolumn}
\usepackage{bm}
\usepackage{color}
\usepackage{comment}
\usepackage{xcolor}
\usepackage[cp1251]{inputenc}

\begin{document}

\title{\bf Anisotropic spheres via embedding approach in $\mathcal{R}+\beta\mathcal{R}^{2}$ gravity with matter coupling}

\author{G. Mustafa}
\email{gmustafa3828@gmail.com}\affiliation{Department of Mathematics, Shanghai University,
Shanghai, 200444, Shanghai, People's Republic of China}

\author{Xia Tie-Cheng}
\email{xiatc@shu.edu.cn}\affiliation{Department of Mathematics, Shanghai University,
Shanghai, 200444, Shanghai, People's Republic of China.}

\author{Mushtaq Ahmad}
\email{mushtaq.sial@nu.edu.pk}\affiliation{National University of Computer and
Emerging Sciences,\\ Chiniot-Faisalabad Campus, Pakistan.}

\author{M. Farasat Shamir}
\email{farasat.shamir@nu.edu.pk; farasat.shamir@gmail.com}\affiliation{National University of Computer and
Emerging Sciences,\\ Lahore Campus, Pakistan.}

\begin{abstract}
The manifesto of the current article is to investigate the compact anisotropic matter profiles in the
context of one of the modified gravitational theories, known as $f(\mathcal{R}, \mathcal{T})$ gravity,
where $\mathcal{R}$ is a Ricci Scalar and $\mathcal{T}$ is the trace of the energy-momentum tensor.
To achieve the desired goal, we capitalized
on the spherical symmetric space–time and utilized the embedding class-1 solution via Karmarkar's
condition in modeling the matter profiles. To calculate the unidentified constraints, Schwarzschild
exterior solution along with experimental statistics of three different stars LMC X-4, Cen X-3, and
EXO 1785-248 are taken under consideration. For the evaluation of the dynamical equations, a unique
model $f(\mathcal{R}, \mathcal{T})=\mathcal{R}+\beta \mathcal{R}^2+\lambda \mathcal{T}$
 has been considered, with $\beta$ and $\lambda$ being the real constants. Different
physical aspects have been exploited with the help of modified dynamical equations. Conclusively, all
the stars under observations are realistic, stable, and are free from all singularities.
\\\\
\textbf{Keywords}: Anisotropic spheres; $f(\mathcal{R}, \mathcal{T})$ gravity; Compact stars; Embedding Class I.
\end{abstract}

\maketitle

\date{\today}

%%%%%%%%%%%%%%%%%%%%%%%%%%%%%%%%%%%%%%%%%%%%%%%%%%%%%%%%%%%%%%%%%%%%%%%%
%%%%%%%%%%%%%%%        Introduction        %%%%%%%%%%%%%%%%%%%%%%%%%%%%%
%%%%%%%%%%%%%%%%%%%%%%%%%%%%%%%%%%%%%%%%%%%%%%%%%%%%%%%%%%%%%%%%%%%%%%%%
\section{Introduction}
Late time evolution of stellar configurations, triggered by an
immense gravitational pull has been anticipated to a great extent in the field of astrophysics and the modified gravitational
theories. It expedites the examination of diverse attributes regarding the gravitating source by physical phenomena. Baade and
Zwicky \cite{1} forecast the inception of highly dense stellar objects
inaugurating the debate that a supernova might be revolutionized
into a highly dense star. This reality came into existence when
exceptionally magnetized as well as rotating neutrons stars were
detected. Therefore, a fundamental shift regarding normal stars
to compact stars came into existence. By the newly discovered
concept, the normal stars shifted into an extensive range, such as
quark stars, neutron stars, gravastars, dark stars, and finally black
holes. The actuality of the extensive range of these stars led the
researchers to curiosity, regarding the formation of these stars.
The stellar death of a normal star occurs, that is when the nuclear
fusion reactions cease to act and burn all of their nuclear fuel
results in the formation of new compact stars. The newly formed
compact stars are primarily distinguished from the normal stars
in two ways. Since all the fuel has been utilized by the star, hence
the star cannot sustain against the gravitational collapse due to
thermal pressure. Analogous to that, the white dwarf is stabilized
due to strong degenerate electron pressure, while the neutron
star is stabilized due to degenerate neutron pressure. Whereas
black holes are entirely the collapsed remnants, therefore there
is neither a thermal pressure nor a degenerate pressure sufficient
enough to repress the centripetal pull of gravity; as a result,
it leads towards the gravitational singularities and the event
horizon. The formed compact stellar remnants consist of huge
density and relatively small radii in contrast to the normal stars.\\

The intention to investigate the physically stable models, leads us to an analytical approach regarding the Einstein field equations. One of the essential tools is to adopt the embedding class I space-time which transforms a four dimensional manifold into a Euclidean space of higher dimension. The conversion of curved
embedding class space–time into higher dimensional space–time is substantial to develop exact new models in the field of astrophysics. The class I embedding condition leads towards a differential equation in the framework of spherically symmetric space–time which connects the gravitational potentials i.e., $g_{rr}$ and $g_{tt}$, the condition is also recognized as the Karmarkar condition \cite{2}. The Karmarkar’s condition appears to be very influential in exploring new solutions for the astrophysical models. Schlia \cite{3} was the pioneer in developing the Karmarkar condition for a spherically symmetric space–time. The embedding theorem based on the isometrics has been presented by Nash \cite{4}. Maurya et al. \cite{5}-\cite{12} were the first explorers in the aspect of applying the embedding approach to the anisotropic matter configurations. After the new dawn of general relativity (GR), $f(\mathcal{R})$ theory is considered to be quite a fascinating tool for the amplification of GR. Further, many researchers presented different versions of this theory, which were also very prosperous in diverse fields. The recent extension of this theory is regarded as $f(\mathcal{R}, \mathcal{T})$ gravity, which was presented by Harko et al. \cite{13}. The $f(\mathcal{R}, \mathcal{T})$ theory has been the center of attention by many analysts and consequently many intriguing cosmological aspects have been unraveled \cite{14}-\cite{17}. The analysis regarding isotropic matter profile of the self-gravitating system and its stability has been done by Sharif et al. \cite{18}. Alhamzawi and Alhamzawi \cite{19} construed the occurrence of lensing of gravitation in the context  of a modified
$f(\mathcal{R}, \mathcal{T})$ theory. Moraes et al. \cite{20} numerically investigated the stability of the gravitational lensing by utilizing the Tolman–Oppenheimer–Volkov (TOV) equations in $f(\mathcal{R}, \mathcal{T})$ gravity. Das et al. \cite{21} formulated a family of solutions by characterizing the interior geometry of compact stars, permitting
conformal motion under the influence of $f(\mathcal{R}, \mathcal{T})$ gravity. Moraes et al. \cite{22} investigated the configurations consisting of hydrostatic equilibrium along with fluids whose pressure was computed from equation of state (EoS) in the light of $f(\mathcal{R}, \mathcal{T})$ gravitational theory. Yousaf et al. \cite{23} investigated the formation of relativistic stellar profiles in the regime of $f(\mathcal{R}, \mathcal{T})$ gravity by utilizing the Krori and Barura model. The study of dense anisotropic profiles consisting
of charge has been investigated by Maurya and Aurtiz \cite{24}. In this regard, they utilized the Durgapal–Fuloria model in the context of $f(\mathcal{R}, \mathcal{T})$ gravity and applied gravitational decoupling utilizing geometric deformation. Waheed et al. \cite{25} analyzed the existence of highly dense stellar configurations by utilizing Karmarkar along with the Pandey–Sharma condition. To do so, they used spherically symmetric space–time in the context of $f(\mathcal{R}, \mathcal{T})$ gravity. Mustafa et al. \cite{26} analyzed the Class 1 embedding condition in the presence of anisotropy matter profile and utilized the interior geometry of Schwarzschild along with Kohler–Chao solutions in modified gravity. The matter configuration consisting of nuclear density of $10^{15}$gm/cc exhibits the behavior of anisotropy i.e. $p_t-p_r\neq0$  which exists due to certain factors involving magnetic flux, viscosity, phase transition, etc. In this regard, Ruderman \cite{27} is the pioneer who argued about the anisotropy existing at the interior of the stars.\\

Modified gravitational theories have provided an overwhelming approach in analyzing the anisotropic stellar configurations
inheriting high matter profiles \cite{28}-\cite{33}. This work aims to analyze the modified $f(\mathcal{R}, \mathcal{T})$ gravity to devise a realistic configuration which in nature is anisotropic. For this purpose, we take into account three different matter profiles i.e. LMC X-4, Cen X - 3,
and EXO 1785–248, and apply a well-known embedding class 1 approach. Moreover, the structural aspect of anisotropic profiles has been examined by making use of spherically symmetric space–time along with the categorical $f(\mathcal{R}, \mathcal{T})$ gravity model. The layout of this article is as follows: In Section 2, modified field equations have been formulated by utilizing the Karmarkar condition. Section 3 is to provide the matching conditions by considering Schwarzschild’s solution. The physical investigation has been done comprehensively in Section 4. Conclusive remarks have been provided in the last Section.
%%%%%%%%%%%%%%%%%%%%%%%%%%%%%%%%%%%%%%%%%%%%%%%%%%%%%%%%%%%%%%%%%%%%%%%%
%%%%%%%%%%%%%%%            Gravity         %%%%%%%%%%%%%%%%%%%%%%%%%%%%%
%%%%%%%%%%%%%%%%%%%%%%%%%%%%%%%%%%%%%%%%%%%%%%%%%%%%%%%%%%%%%%%%%%%%%%%%
\section{$f(\mathcal{R},\mathcal{T})$ Theory of Gravity }

The modified form of Einstein-Hilbert action for extended $f(\mathcal{R},\mathcal{T})$ theory of gravity is defined as follows:
\begin{equation}\label{1}
\mathcal{S}=\frac{1}{2}\int \left[L_{\mathrm{m}}+f(\mathcal{R}, \mathcal{T})\right]\sqrt{-g}d^{4}x,
\end{equation}
where $L_{\mathrm{m}}$ and $f(\mathcal{R},\mathcal{T})$ denote the matter Lagrangian density and
a modified function, respectively. Here, $\mathcal{R}$ and $\mathcal{T}$ are known as scalar curvature and trace
of the energy-momentum tensor, respectively. Now, by varying Eq.(\ref{1}), we get the following modified
set of equations
\begin{equation}\label{2}
(1-f_{\mathcal{T}}(\mathcal{R},\mathcal{T}))\mathcal{T}_{\mu\nu}-f_{\mathcal{T}}(\mathcal{R},\mathcal{T})\Theta_{\mu\nu}=-\frac{1}{2}f(\mathcal{R},\mathcal{T})g_{\mu\nu}
+(\mathcal{R}_{\mu\nu}+(g_{\mu\nu}\Box-\nabla_{\mu}\nabla_{\nu}))f_{\mathcal{R}}(\mathcal{R}
,\mathcal{T}),
\end{equation}
where,
\begin{eqnarray}
\Theta_{\mu\nu}&=&\frac{g^{\alpha\beta}\delta \mathcal{T}_{\mu\nu}}{\delta g^{\mu\nu}}=-2g^{\alpha\beta}\frac{\partial^{2}L_{\mathrm{m}}}{\partial g^{\mu\nu}\partial g^{\alpha\beta}}-2\mathcal{T}_{\mu\nu}+g_{\mu\nu}L_{\mathrm{m}},\nonumber\\
\Box&=&\frac{\partial_{\mu}(\sqrt{-g}g^{\mu\nu}\partial_{\nu})}{\sqrt{-g}},\quad
f_{\mathcal{R}}(\mathcal{R},\mathcal{T})=\frac{\partial f(\mathcal{R},\mathcal{T})}{\partial \mathcal{R}},\quad
f_{T}(\mathcal{R},\mathcal{T})=\frac{\partial f(\mathcal{R},\mathcal{T})}{\partial \mathcal{T}},\nonumber
\end{eqnarray}
with $\nabla$, representing the covariant derivative. The energy-momentum tensor with the anisotropic
matter source is defined as
\begin{equation}\label{3}
\mathcal{T}_{\mu\nu}=\rho U_{\mu}U_{\nu}+p_{r}V_{\mu}V_{\nu}+p_{t}(U_{\mu}
U_{\nu}-g_{\mu\nu}-V_{\mu}V_{\nu}),
\end{equation}
where $U_{\mu}$ represents the vector for 4-velocity and $V_{\mu}$ is a vector in the direction of
radial pressure. Further, the expressions, i.e., $\rho,~p_{t}$ and $p_{r}$ are used to define define energy density, tangential and radial
components of pressure, respectively.

Using Eq.(\ref{3}) in Eq.(\ref{2}), we get the following set of equation:
\begin{eqnarray}\label{4}
\mathcal{G}_{\mu\nu}&=&\frac{1}{f_{\mathcal{R}}(\mathcal{R},\mathcal{T})}\left((1+f_{\mathcal{T}}(\mathcal{R},\mathcal{T}))\mathcal{T}_{\mu\nu}+(\nabla_{\mu}\nabla_{\nu}-g_{\mu\nu}\Box)
f_{\mathcal{R}}(\mathcal{R},\mathcal{T})+
\frac{1}{2}(f(\mathcal{R},\mathcal{T})-\mathcal{R} f_{\mathcal{R}}(\mathcal{R},\mathcal{T}))g_{\mu\nu}\right.\nonumber\\&-&\left.\rho g_{\mu\nu}f_{\mathcal{T}}(\mathcal{R},\mathcal{T})\right).
\end{eqnarray}
We assume a static and spherically symmetric line element,
which is defined as:
\begin{equation}\label{5}
 d{s}^2=-e^{a(r)} dt^2+e^{b(r)} d{r}^2+r^2d\Omega^{2},
\end{equation}
where the expression $d\Omega^{2}$ defines the $g_{\theta\theta}=r^2d\theta^{2}$ and $g_{\phi\phi}=r^2\sin^{2}\theta d\phi^{2}$
components, $e^{b(r)}$ and $e^{a(r)}$ denote the gravitational components of stellar geometry. Further, we fix a quadratic model of $f(\mathcal{R},\mathcal{T})$, which is defined as:
\begin{equation}\label{6}
f(\mathcal{R},\mathcal{T})=\mathcal{R}+\beta\times\mathcal{R}^{2}+\lambda\times\mathcal{T}.
\end{equation}
The considered model $f(\mathcal{R}, \mathcal{T}) = \mathcal{R} + \beta \mathcal{R}^2 + \lambda \mathcal{T}$ involves a particular case of well-known Starobinsky model \cite{34} with matter coupling. It is an interesting point that, in the Starobinsky model $\mathcal{R} + \beta \mathcal{R}^2$, a maximum value of $M/M_{\odot}$ or beyond is reached when the value of the parameter $\beta$ is selected to be negative. But, this leads to an issue; specifically, the Ricci scalar performs a damped oscillation. On the other hand, the Ricci scalar smoothly decreases to zero as we approach towards infinity for positive values of parameter $\beta$, for which the star can support a maximum mass lower than  $2M/M_{\odot}$. Now, we elaborate an eminent Karmarkar condition concisely which is the integral tool for current study. The infrastructure connecting the Karmarkar condition is established on the class I space of Riemannian geometry. A sufficient condition comprises of a second order symmetric tensor and the Riemann Christoffel tensor, given as
\begin{equation*}
\Sigma(\Lambda_{\mu\eta}\Lambda_{\upsilon\gamma} - \Lambda_{\mu\gamma}\Lambda_{\nu\eta})=\mathcal{R}_{\mu\upsilon\eta\gamma},
\end{equation*}
\begin{equation*}
\Lambda_{\mu\nu};n-\Lambda_{\nu\eta};\nu=0.
\end{equation*}
Here ; stands for covariant derivative whereas $\Sigma=\pm1$. These
values signify a space-like or time-like manifold relying on the
sign considered as $-$ or $+$. Now, the Karmarkar condition is
defined as
\begin{equation}\label{7}
\mathcal{R}_{1414}\mathcal{R}_{2323}=\mathcal{R}_{1224}\mathcal{R}_{1334}+ \mathcal{R}_{1212}\mathcal{R}_{3434}.
\end{equation}
These Riemann tensor components are given below as follows.
\begin{eqnarray*}
\mathcal{R}_{1414}&=&\frac{e^{a (r)}(2a ''(r)+a '(r)^2-a '(r)b'(r))}{4},\;\;\;\;\;\mathcal{R}_{2323}=\frac{r^{2}sin^{2}\theta (e^{b (r)}-1)}{e^{b (r)}},\nonumber\\
\mathcal{R}_{1212}&=&\frac{r b '(r)}{2},\;\;\;\;\;\;\;\;\;\;\;\;\;\;\;\;\;\;\;\;\;\;\;\;\;\;\;\;\;\;\;\;\;\;\;\;\;\;\;\;\;\;\;\;\;\;\;\;\;\mathcal{R}_{3434}=\frac{rsin^{2}\theta b '(r) e^{a (r)}-b (r)}{2},\nonumber\\
\mathcal{R}_{1334}&=&\mathcal{R}_{1224}sin^{2}\theta,\;\;\;\;\;\;\;\;\;\;\;\;\;\;\;\;\;\;\;\;\;\;\;\;\;\;\;\;\;\;\;\;\;\;\;\;\;\;\;\;\;\;\mathcal{R}_{1224}= 0,\nonumber
\end{eqnarray*}
where, $\mathcal{R}_{2323}\neq0$. A differential equation can be achieved by
utilizing the Karmarkar condition using Eq. (\ref{7}) as
\begin{equation}\label{8}
\frac{a'(r) b'(r)}{1-e^{b(r)}}-\left(a'(r) b'(r)+a'(r)^2-2 \left(a''(r)+a'(r)^2\right)\right)=0,\;\;\;\;\;\;\;e^{b(r)}\neq1.
\end{equation}
Integration of Eq. (\ref{8}) provides a connection between two main
gravitational components of the space-time as follows
\begin{equation}\label{9}
e^{b(r)}=e^{a(r)}\times a'(r)^2+1+K,
\end{equation}
where $K$ is a constant of integration. We choose a specific model for a $g_{tt}$ component which is expressed as
\begin{equation}\label{10}
e^{a(r)}=\psi_{1} \left(r^{2} \psi_{2}+1\right)^{n},
\end{equation}
where $\psi _1$, $\psi _2$ are assumed as constants, $n$ is an integer. By plugging Eq. (\ref{10}) in Eq. (\ref{9}), we get the $g_{rr}$ component, which is calculated as
\begin{equation}\label{11}
e^{b(r)}=r^2 \psi _2 \psi _3 \left(r^2 \psi _2+1\right){}^{n-2}+1,
\end{equation}
where $\psi_{3}=4 n^{2}\times \psi_{2}\times \psi_{1}\times K$. It is mentioned here that we
get realistic results for $n > 2$. Now, we are able to calculate the
following set of modified field equations for the anisotropic stellar
configuration.
\begin{eqnarray}
\rho&=&\frac{1}{2 (\lambda +1) (2 \lambda +1) \left(r^3 \psi _2 \psi _3 \varUpsilon _1^{n-2}+r\right)^2}\bigg(\frac{\psi _2}{\varUpsilon _1^{3-n}+r^2 \psi _2 \psi _3 \varUpsilon _1}\times\bigg(-\frac{12 \beta  (\lambda +1) n^2 r^2 \psi _2 \left(-\varUpsilon _1\right){}^2}{\varUpsilon _1^4}+\varUpsilon _2\nonumber\\ &+&\frac{4 \beta  (3 \lambda +1) n^4 r^6 \psi _2^3}{\varUpsilon _1^4}-\frac{8\beta  (3 \lambda +2)}{\varUpsilon _1^4}\times\bigg(-\frac{\psi _3 \varUpsilon _{16} \varUpsilon _1^{n+2}}{\varUpsilon _5^2}+\psi _2r^2\left(4 n \varUpsilon _1 \left(r^2 \psi _2-3\right)-3 n \left(r^4 \psi _2^2-6 r^2 \psi _2+1\right)\right.\nonumber\\&-&\left.\right.\frac{2 \psi _3 \varUpsilon _1^{n+1}}{\varUpsilon _5^3}\times\left(r^2 \psi _2 \psi _3^2 \left(\varUpsilon _1 \varUpsilon _8\right) \varUpsilon _1^{2 n}-\psi _3 \varUpsilon _7 \varUpsilon _1^{n+2}+(n-2) \varUpsilon _6 \varUpsilon _1^4\right)-\frac{2 n r^2 \left(-\varUpsilon _1\right) \varUpsilon _{10}}{\varUpsilon _1^2}-\varUpsilon _9-\varUpsilon _{11}+\varUpsilon _{12}\bigg)\nonumber\\&+&\frac{n^2 r^2 \psi _2 \varUpsilon _{13}}{\varUpsilon _1^2}+2 \psi _3 \varUpsilon _1^{n-2} \left(2 \beta  (9 \lambda +5)+\varUpsilon _{14} \left((2 \lambda +1) r^2-2 (3 \beta  \lambda +\beta )\right)\right)-\frac{n}{\varUpsilon _1}\times\bigg(\lambda  \varUpsilon _{14} \left(\frac{2 r^4 \psi _3 \psi _2 \varUpsilon _{15}}{\varUpsilon _1^{3-n}+r^2 \psi _2 \psi _3 \varUpsilon _1}\right.\nonumber\\&-&\left.4 \left(6 \beta +r^2\right)\right)+\beta  \left(24 \lambda +2 r^2 \psi _2 \left(\frac{\varUpsilon _{19}}{\varUpsilon _1^{3-n}+r^2 \psi _2 \psi _3 \varUpsilon _1}+\varUpsilon _{17}-\varUpsilon _{18}\right)\right)\bigg)\bigg)\bigg),\label{19}\\
p_{r}&=&\frac{-1}{4 (\lambda +1) (2 \lambda +1) r^3 \varUpsilon _{14}^2}\bigg(-4 \beta  r \psi _2 \left(\frac{2 \psi _3 \varUpsilon _{15} \varUpsilon _{24}}{\varUpsilon _1^{3-n}+r^2 \psi _2 \psi _3 \varUpsilon _1}+\frac{4 r^2 \psi _2 \varUpsilon _{23}}{\varUpsilon _1^4}+\varUpsilon _{21}+\varUpsilon _{22}\right)-2 \varUpsilon _{14} \varUpsilon _{20}+4 r \psi _3\nonumber\\&\times& \psi _2 \varUpsilon _1^{n-2}\frac{2 n^2 r^3 \psi _2^2 \varUpsilon _{25}}{\varUpsilon _1^2}+\frac{2 n r \psi _2}{\varUpsilon _1}+\left(\varUpsilon _{14} \left((2 \lambda +1) r^2-2 (3 \beta  \lambda +\beta )\right)-2 \beta  (15 \lambda +7)\right)\bigg(\beta  \left(8 (5 \lambda +4)+2 r^2 \psi _2 \right.\nonumber\\&\times&\left.\left(\frac{8 (\lambda +2) \psi _3 \varUpsilon _{16} \varUpsilon _1^{n-2}}{\varUpsilon _5^2}+\frac{4 (5 \lambda +6) n \left(-\varUpsilon _1\right)}{\varUpsilon _1^2}+\frac{\psi _3 \varUpsilon _{15} \varUpsilon _{28}}{\varUpsilon _1^{3-n}+r^2 \psi _2 \psi _3 \varUpsilon _1}+\varUpsilon _{26}-\varUpsilon _{27}\right)\right)-\varUpsilon _{14}  \nonumber\\&\times&\left(\frac{2 \lambda  r^4 \psi _2 \psi _3 \varUpsilon _{15}}{\varUpsilon _1^{3-n}+r^2 \psi _2 \psi _3 \varUpsilon _1}+4 \left((\lambda +1) r^2-6 \beta  \lambda \right)\right)\bigg)\bigg),\label{19}\\
p_{t}&=&\frac{\psi _2}{2 (\lambda +1) (2 \lambda +1) \left(r^3 \psi _2 \psi _3 \varUpsilon _1^{n-2}+r\right)^2}\bigg(\frac{4 \beta  \varUpsilon _{31}}{\varUpsilon _1^4}+\frac{n^2 r^2 \psi _2 \varUpsilon _{33}}{\varUpsilon _1^2}+\frac{2 \psi _3 \varUpsilon _{15} \varUpsilon _{32}}{\varUpsilon _1^{3-n}+r^2 \psi _2 \psi _3 \varUpsilon _1}-\frac{1}{\varUpsilon _1}\times n \left(\beta  \left(-8 \lambda\right.\right. \nonumber\\&+&\left.\left.2 r^2 \psi _2 \left(\frac{4 (8 \lambda +7) \psi _3 \varUpsilon _{16} \varUpsilon _1^{n-2}}{\varUpsilon _5^2}+\frac{4 (3 \lambda +5) n \left(-\varUpsilon _1\right)}{\varUpsilon _1^2}-\frac{4 (7 \lambda +6) n r^2 \psi _2 \left(r^2 \psi _2-3\right)}{\varUpsilon _1^3}-\frac{4 (24 \lambda +19) r^2 \psi _3^2 \psi _2 \varUpsilon _{15}^2}{\left(\varUpsilon _1^{3-n}+r^2 \psi _2 \psi _3 \varUpsilon _1\right){}^2}\right.\right.\right. \nonumber\\&+&\left.\left.\left.\frac{\varUpsilon _{36}}{\varUpsilon _1^{3-n}+r^2 \psi _2 \psi _3 \varUpsilon _1}+\frac{12 (3 \lambda +2) r^4 \psi _2^2 \psi _3^3 \varUpsilon _{15}^3}{\left(\varUpsilon _1^{3-n}+r^2 \psi _2 \psi _3 \varUpsilon _1\right){}^3}+\frac{\varUpsilon _{35}}{\varUpsilon _5^3}\right)+8\right)+\varUpsilon _{14} \varUpsilon _{34}\right)+\varUpsilon _{29}-\varUpsilon _{30}\bigg),
\end{eqnarray}
where $\varUpsilon _i$, $\{i=1,...,36\}$, are given in the Appendix (\textbf{I}).

\section{Comparison of exterior and interior solution}

By considering the Jebsen-Birkhoff's theorem, the spherically
symmetric vacuum solution of GR field equations must be asymptotically flat. In particular, the spacetime is of the form
\begin{equation}\label{14}
 d{s}^2=\varepsilon dt^2-\varepsilon^{-1} d{r}^2-r^2(d\theta^{2}+sin^{2}\theta d\phi^{2}),
\end{equation}
where $\varepsilon=\left(1-\frac{2 M}{r}\right)$. Here, $M$ denotes the stellar mass of the star. Now, considering the constraint $p_{r}(r = R_{\epsilon})=0$ at the boundary $r=R_{\epsilon}$ and using metric coefficients $g_{tt}$ and $g_{rr}$ from Eq. (\ref{5}) and Eq. (\ref{14}), we calculate the following expressions
\begin{eqnarray}
\psi_{1} \left(\psi_{2} R_{\epsilon}^2+1\right)^n&=&1-\frac{2 M}{R_{\epsilon}},\label{15}\\
\psi_{3} \psi_{2} R_{\epsilon}^2 \left(\psi_{2} R_{\epsilon}^2+1\right)^{n-2}+1&=&\left(1-\frac{2 M}{R_{\epsilon}}\right)^{-1},\label{16}\\
n \psi_{1} \psi_{2} \left(\psi_{2} R_{\epsilon}^2+1\right)^{n-1}&=& \frac{M}{R_{\epsilon}^3},\label{17}\\
p_{r}(r=R_{\epsilon})&=&0.\label{18}
\end{eqnarray}
Utilizing these boundary conditions from Eqs. (\ref{15}-\ref{18}), we get the following relations
\begin{eqnarray}
\psi_{1}&=&\frac{(R_{\epsilon}-2 M) \left(1-\frac{M}{2 n M-n R_{\epsilon}+M}\right)^{-n}}{R_{\epsilon}},\label{19}\\
\psi_{2}&=&\frac{M}{R_{\epsilon}^2 (n R_{\epsilon}-(2 n+1) M)},\label{20}\\
\psi_{3}&=&2 n \left(1-\frac{M}{2 n M-n R_{\epsilon}+M}\right)^{1-n},\label{21}\\
\lambda &=&\frac{\lambda _1+\lambda _2+\lambda _3+\lambda _4}{\lambda _5+\lambda _6+\lambda _7+\lambda _8}\label{22},
\end{eqnarray}
where $\lambda _i$, $\{i=1,...,8\}$ are given in the Appendix (\textbf{II}).\\

The estimated values of the above parameters, i.e., $\psi_{1},\;\psi_{2},\;\psi_{3},\;\&\;\lambda $ are given in Table \textbf{I}, Table \textbf{II} and Table \textbf{III}.
\begin{center}
\begin{table}
\caption{\label{tab1}{Predicted values of $\psi_{1},\;\psi_{2},\;\psi_{3}$, and $\lambda$ with $\beta =2$ and and (Radii=9.711 km, \& Mass =1.29$M/M_{\odot}$).}}
\begin{tabular}{|c|c|c|c|c|c|c|c|c|}
    \hline
    & \multicolumn{4}{|c|}{LMC X-4} \\
    \hline
$n$ \;\;\;\;                               & $\psi_{1}$\;\;\;\       &$\psi_{2}$\;\;\;     &$\psi_{3}$\;\;\;\;\;\; &$\lambda$ \\
\hline
3\;\;\;\;                                &0.4329\;             &0.001273\;\;\;\;   &4.7820\;\;\;\;     &0.066676\\
5\;\;\;\;                                &0.4363\;             &0.000729\;\;\;\;   &7.6640\;\;\;\;     &0.040511\\
10\;\;\;\;                               &0.4387\;             &0.000352\;\;\;\;   &14.900\;\;\;\;     &0.021340\\
20\;\;\;\;                               &0.4399\;             &0.000173\;\;\;\;   &29.392\;\;\;\;     &0.011919\\
50\;\;\;\;                               &0.4406\;             &0.000068\;\;\;\;   &72.882\;\;\;\;     &0.006322\\
100\;\;\;\;                              &0.4408\;             &0.000034\;\;\;\;   &145.373\;\;\;\;    &0.004466\\
500\;\;\;\;                              &0.4410\;      &6.827742$\times10^{-6}$\; &725.285\;\;\;\;    &0.002985\\
\hline
\end{tabular}
\end{table}
\end{center}
\begin{center}
\begin{table}
\caption{\label{tab1}{Predicted values of $\psi_{1},\;\psi_{2},\;\psi_{3}$, and $\lambda$ with $\beta =2$ and (Radii= 10.136 km, \& Mass =1.49$M/M_{\odot}$).}}
\begin{tabular}{|c|c|c|c|c|c|c|c|c|}
    \hline
    & \multicolumn{4}{|c|}{Cen X - 3} \\
    \hline
$n$ \;\;\;\;                               & $\psi_{1}$\;\;\;\       &$\psi_{2}$\;\;\;     &$\psi_{3}$\;\;\;\;\;\; &$\lambda$ \\
\hline
3\;\;\;\;                                &0.3765\;             &0.001421\;\;\;\;   &4.5683\;\;\;\;     &0.055841\\
5\;\;\;\;                                &0.3807\;             &0.000805\;\;\;\;   &7.2751\;\;\;\;     &0.032774\\
10\;\;\;\;                               &0.3838\;             &0.000386\;\;\;\;   &14.082\;\;\;\;     &0.015565\\
20\;\;\;\;                               &0.3852\;             &0.000186\;\;\;\;   &28.267\;\;\;\;     &0.007027\\
50\;\;\;\;                               &0.3861\;             &0.000074\;\;\;\;   &68.656\;\;\;\;     &0.007027\\
100\;\;\;\;                              &0.3864\;             &0.000036\;\;\;\;   &138.251\;\;\;\;    &0.000239\\
500\;\;\;\;                              &0.3866\;      &7.447003$\times10^{-6}$\; &682.743\;\;\;\;    &0.000113\\
\hline
\end{tabular}
\end{table}
\end{center}

\begin{center}
\begin{table}[h]
\caption{\label{tab1}{Predicted values of $\psi_{1},\;\psi_{2},\;\psi_{3}$, and $\lambda$ with $\beta =2$ and (Radii=8.849 km, \& Mass =1.30$M/M_{\odot}$).}}
\begin{tabular}{|c|c|c|c|c|c|c|c|c|}
    \hline
    & \multicolumn{4}{|c|}{EXO 1785-248} \\
    \hline
$n$ \;\;\;\;                               & $\psi_{1}$\;\;\;\       &$\psi_{2}$\;\;\;     &$\psi_{3}$\;\;\;\;\;\; &$\lambda$ \\
\hline
3\;\;\;\;                                &0.3768\;             &0.001862\;\;\;\;   &4.5698\;\;\;\;     &0.076448\\
5\;\;\;\;                                &0.3811\;             &0.001055\;\;\;\;   &7.2777\;\;\;\;     &0.044225\\
10\;\;\;\;                               &0.3842\;             &0.000496\;\;\;\;   &14.361 \;\;\;\;    &0.020799\\
20\;\;\;\;                               &0.3856\;             &0.000248\;\;\;\;   &27.733\;\;\;\;     &0.009361\\
50\;\;\;\;                               &0.3864\;             &0.000098\;\;\;\;   &68.685\;\;\;\;     &0.002592\\
100\;\;\;\;                              &0.3867\;             &0.000048\;\;\;\;   &136.944\;\;\;\;    &0.000352\\
500\;\;\;\;                              &0.3869\;      &9.759977$\times10^{-6}$\; &683.029\;\;\;\;    &0.000134\\
\hline
\end{tabular}
\end{table}
\end{center}

\section{Physical Analysis}

In this section, we briefly present the results by analyzing the physical attributes along with different aspects of the stellar
configurations under the acquired $f(\mathcal{R}, \mathcal{T})$ gravity model. In order to fulfill the purpose, experimental data of distinct stars i.e. LMC
X-4, Cen X – 3 and EXO 1785–248 is used. All the attributes of the stellar configurations are depicted in tabular form as well as graphically.
\subsection{Gravitational Metric Potential}
\begin{figure}[h]
\centering \epsfig{file=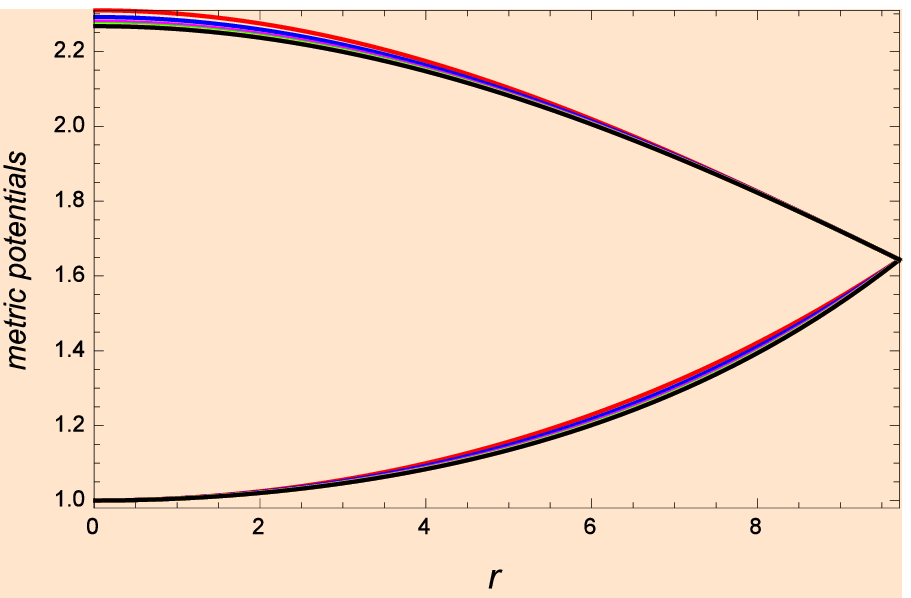, width=.32\linewidth,
height=2in}\epsfig{file=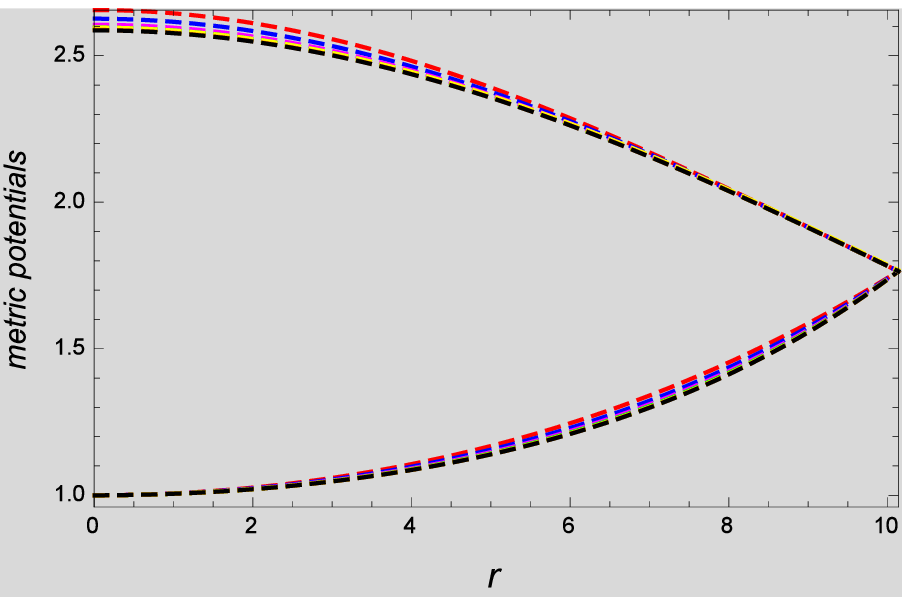, width=.32\linewidth,
height=2in}\epsfig{file=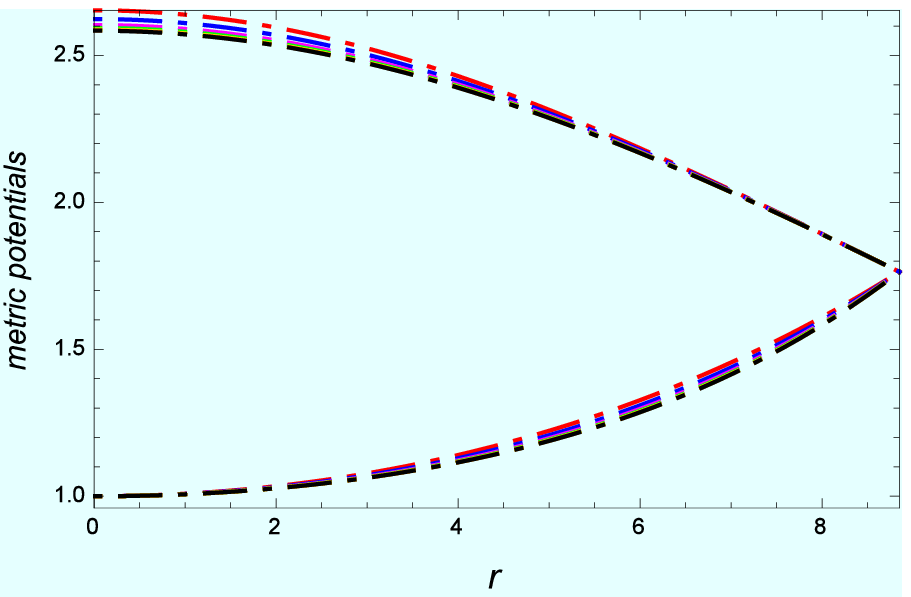, width=.32\linewidth,
height=2in}\caption{ Visual representation of gravitational potentials with $n=3(\textcolor{red}{\bigstar})$, $n=5(\textcolor{blue}{\bigstar})$, $n=10(\textcolor{magenta}{\bigstar})$, $n=20(\textcolor{green}{\bigstar})$, $n=50(\textcolor{yellow}{\bigstar})$, $n=100(\textcolor{orange}{\bigstar})$, and $n=500(\textcolor{black}{\bigstar})$}
\label{Fig.1}
\end{figure}

The existence of anomalies within the sphere such as geometric singularities are contemplated to be an essential peculiarity
in the investigation of stellar spheres. In order to unravel the
existence of singularities, we examine the nature of gravitational
potential $g_{tt}=e^{a}$ and $g_{rr}=e^{b}$ at the core $r=0$ of the
sphere. Physical essence and endurance of the models rely upon
the gravitational metric potentials and it should be decreasing
on regular intervals within the spherical structures. It can be
observed from Fig. \ref{Fig.1} that the metric potential with in the interior
of the sphere exhibits the behavior of $e^{b(r=0)}=1$  and $e^{a(r=0)}\neq 0$
which is consistent and physically valid. It can also be observed
that both of the metric potentials exhibit minimum values at
the center and show non-linear increasing behavior towards the
boundary.
\subsection{Energy Density and Pressure Evolutions}

\begin{figure}[h]
\centering \epsfig{file=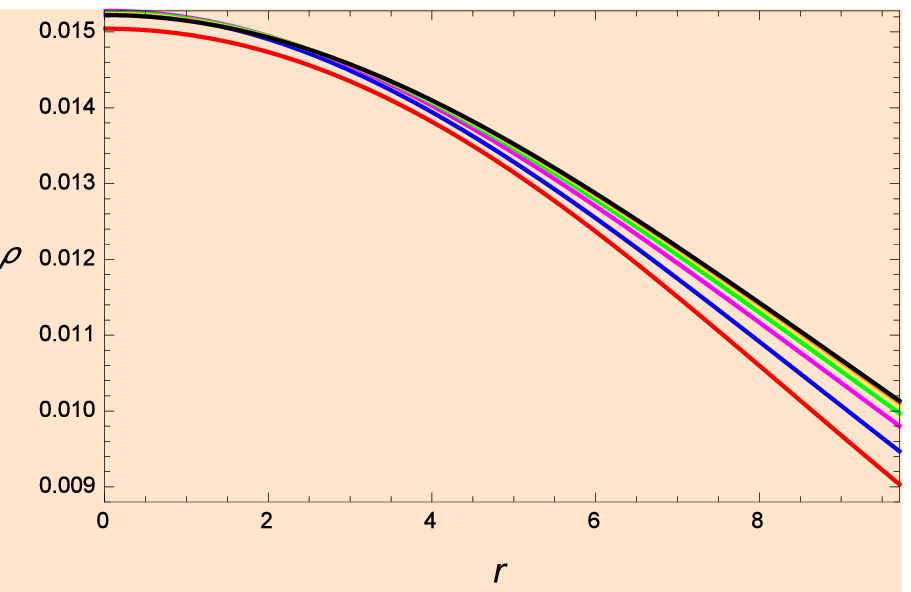, width=.32\linewidth,
height=2in}\epsfig{file=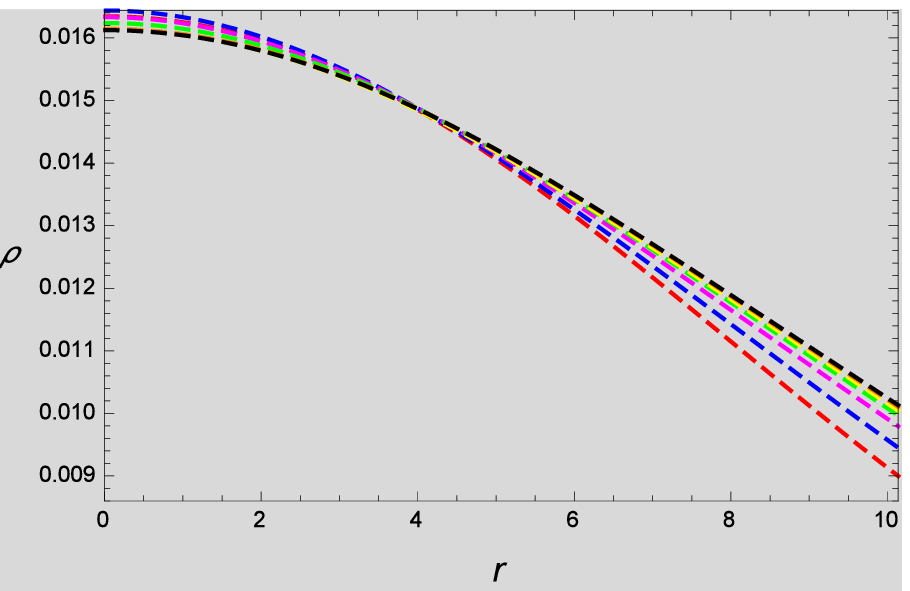, width=.32\linewidth,
height=2in}\epsfig{file=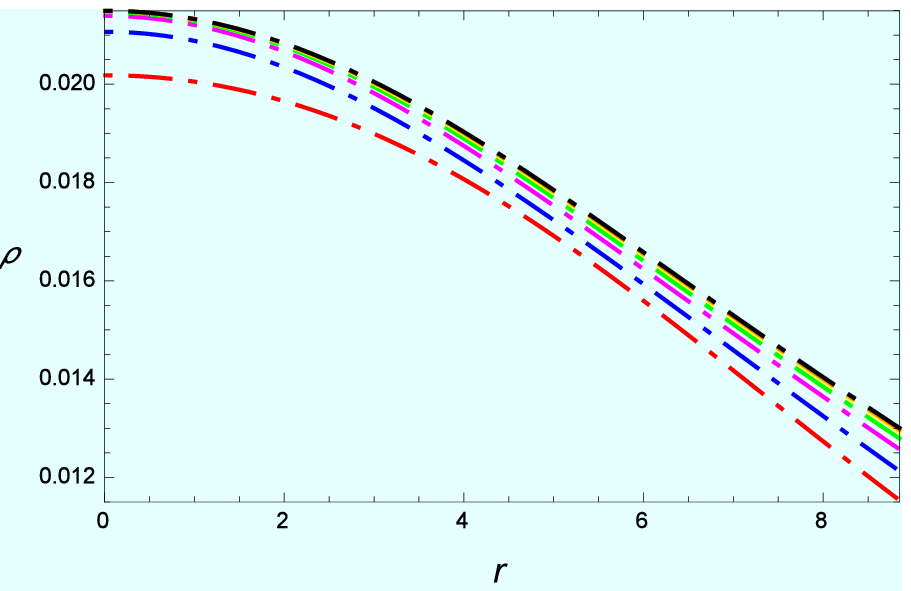, width=.32\linewidth,
height=2in}\caption{Visual representation of $\rho$ with $n=3(\textcolor{red}{\bigstar})$, $n=5(\textcolor{blue}{\bigstar})$, $n=10(\textcolor{magenta}{\bigstar})$, $n=20(\textcolor{green}{\bigstar})$, $n=50(\textcolor{yellow}{\bigstar})$, $n=100(\textcolor{orange}{\bigstar})$, and $n=500(\textcolor{black}{\bigstar})$}
\label{Fig.2}
\end{figure}

\begin{figure}[h]
\centering \epsfig{file=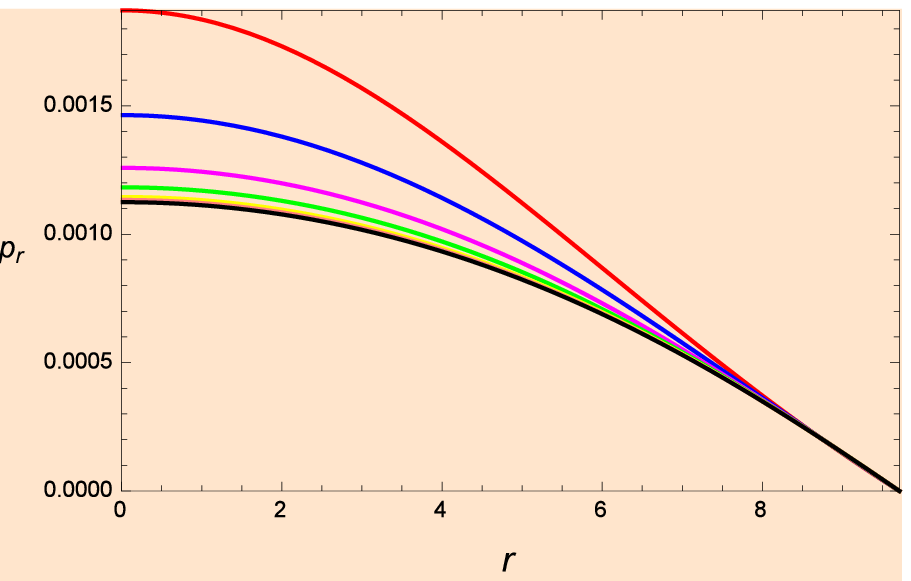, width=.32\linewidth,
height=2in}\epsfig{file=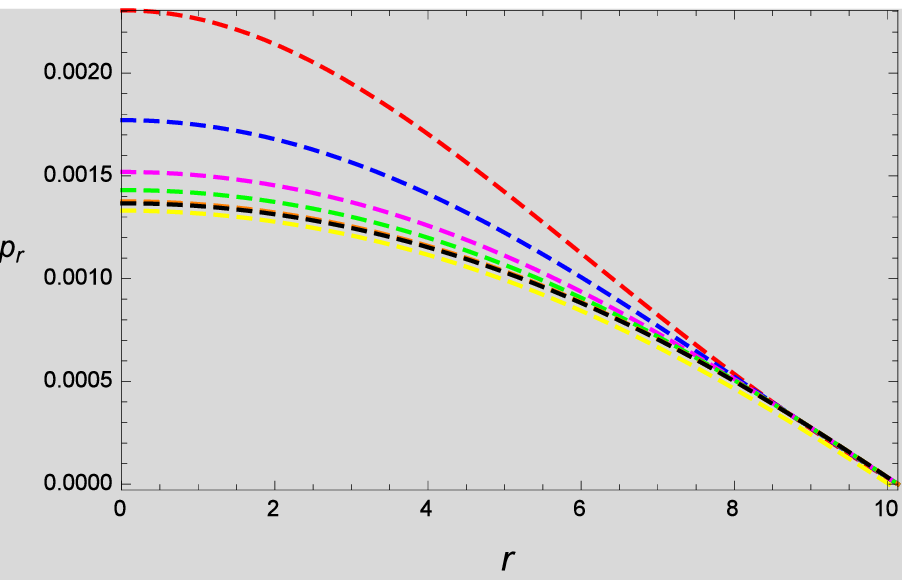, width=.32\linewidth,
height=2in}\epsfig{file=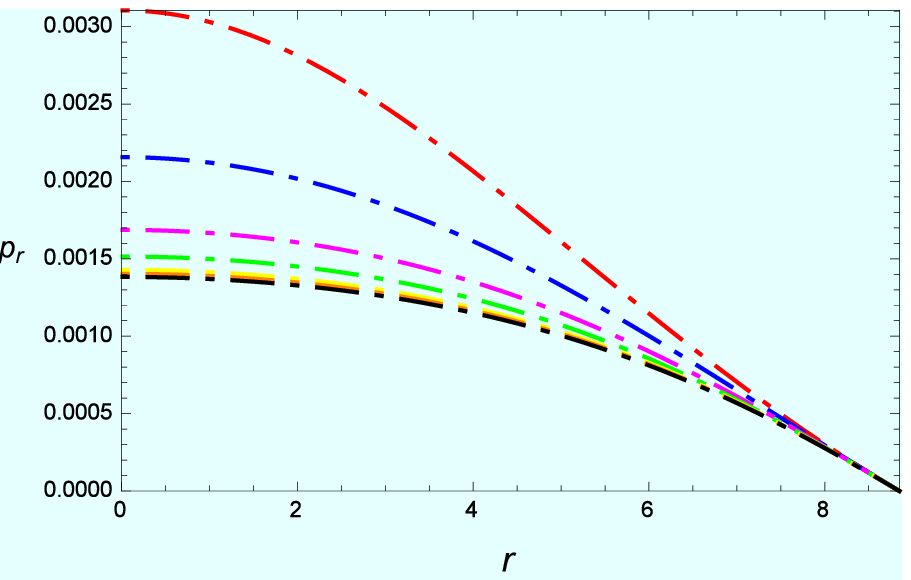, width=.32\linewidth,
height=2in}\caption{Visual representation of $p_r$ with $n=3(\textcolor{red}{\bigstar})$, $n=5(\textcolor{blue}{\bigstar})$, $n=10(\textcolor{magenta}{\bigstar})$, $n=20(\textcolor{green}{\bigstar})$, $n=50(\textcolor{yellow}{\bigstar})$, $n=100(\textcolor{orange}{\bigstar})$, and $n=500(\textcolor{black}{\bigstar})$}
\label{Fig.3}
\end{figure}

\begin{figure}[h]
\centering \epsfig{file=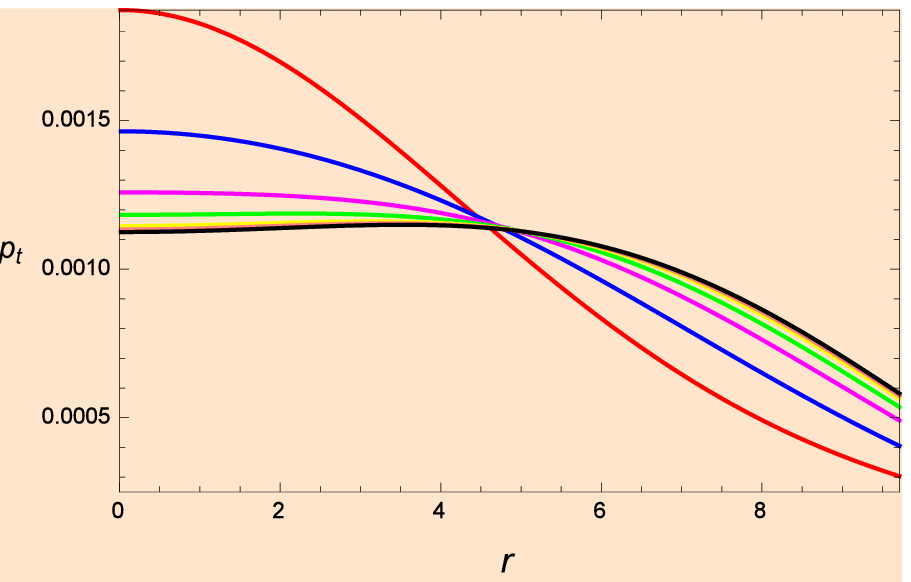, width=.32\linewidth,
height=2in}\epsfig{file=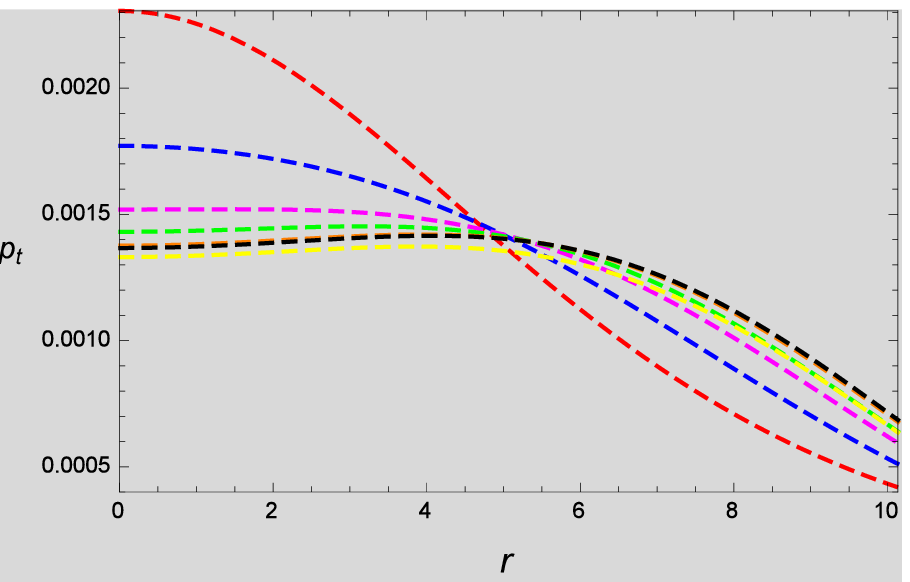, width=.32\linewidth,
height=2in}\epsfig{file=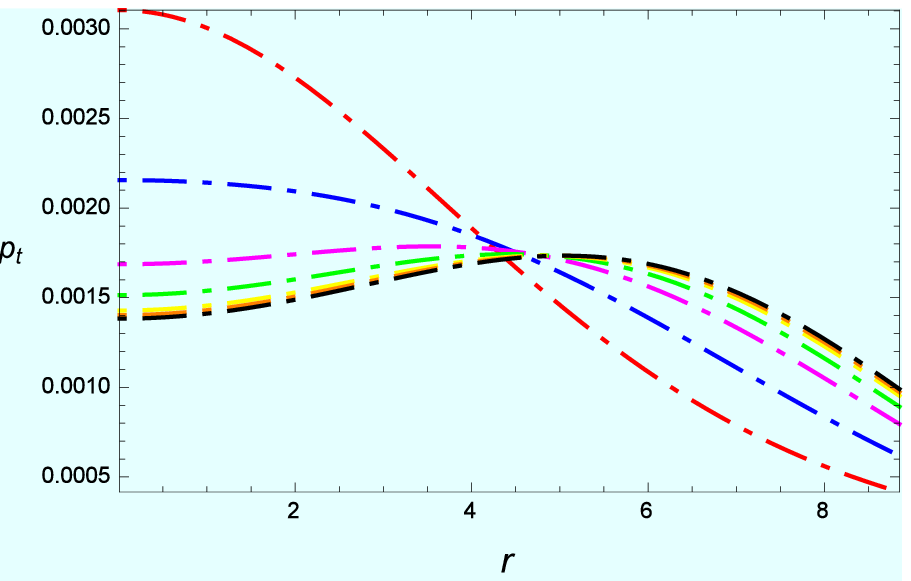, width=.32\linewidth,
height=2in}\caption{Visual representation of $p_t$ with $n=3(\textcolor{red}{\bigstar})$, $n=5(\textcolor{blue}{\bigstar})$, $n=10(\textcolor{magenta}{\bigstar})$, $n=20(\textcolor{green}{\bigstar})$, $n=50(\textcolor{yellow}{\bigstar})$, $n=100(\textcolor{orange}{\bigstar})$, and $n=500(\textcolor{black}{\bigstar})$}
\label{Fig.4}
\end{figure}

Prior to analysis of the anisotropy, we investigate the evolutional change of the matter profiles in connection to the energy
density $\rho$ along with anisotropic stresses such as $p_r$ and $p_t$. The
energy density along with pr and pt exhibits the exceptional
behavior of high density of matter configuration. The phenomenal
high density is due to the strong forces of attraction which are
regarded as dipole interactions and intermolecular forces. Numerical values of density and the components of the pressure for
the three compact spheres are provided in Tables \textbf{IV}-\textbf{VI}. All of the
physical attributes remain positive and appear to be finite
at the core. It confirms that the current system is independent of
all singularities. From the Figs. \ref{Fig.2}-\ref{Fig.4}, it is evident that the matter
configuration under consideration attains the maximum mass at
the core and tends to zero at the boundary of the star, which depicts the high compactness of the stellar spheres. These graphical
plots establish the presence of anisotropy of the compact sphere
under the influence of our $f(\mathcal{R}, \mathcal{T})$ model.\\

\subsection{Anisotropy and Gradients}
\begin{figure}[h]
\centering \epsfig{file=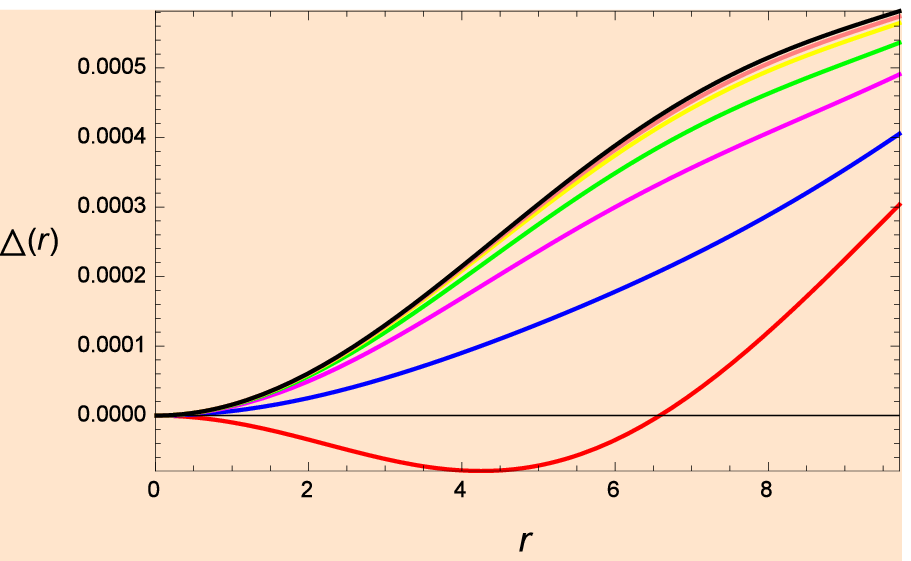, width=.32\linewidth,
height=2in}\epsfig{file=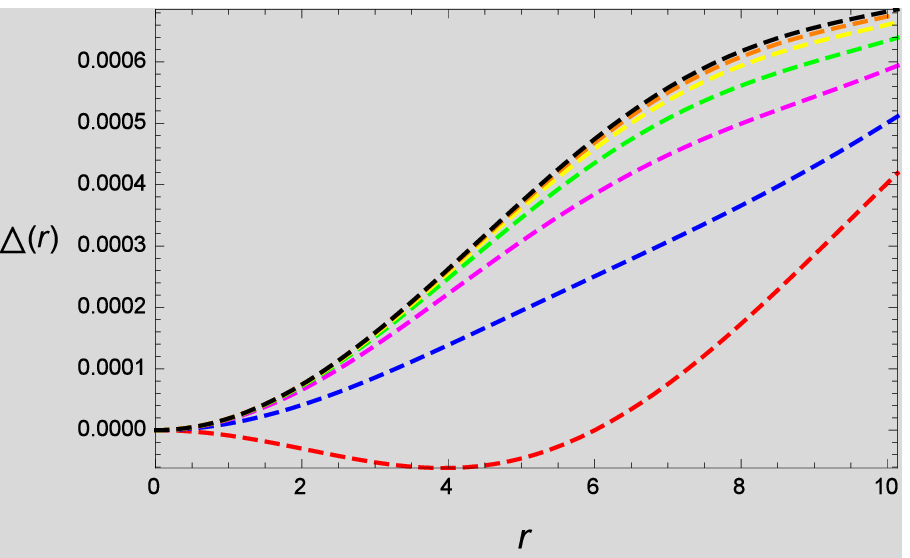, width=.32\linewidth,
height=2in}\epsfig{file=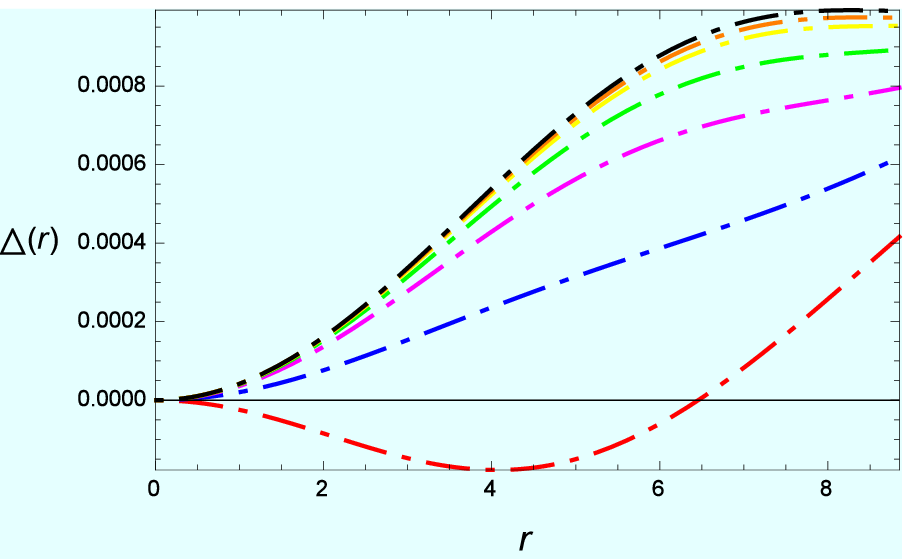, width=.32\linewidth,
height=2in}\caption{Visual representation of $\Delta$ with $n=3(\textcolor{red}{\bigstar})$, $n=5(\textcolor{blue}{\bigstar})$, $n=10(\textcolor{magenta}{\bigstar})$, $n=20(\textcolor{green}{\bigstar})$, $n=50(\textcolor{yellow}{\bigstar})$, $n=100(\textcolor{orange}{\bigstar})$, and $n=500(\textcolor{black}{\bigstar})$}
\label{Fig.5}
\end{figure}

\begin{figure}[h]
\centering \epsfig{file=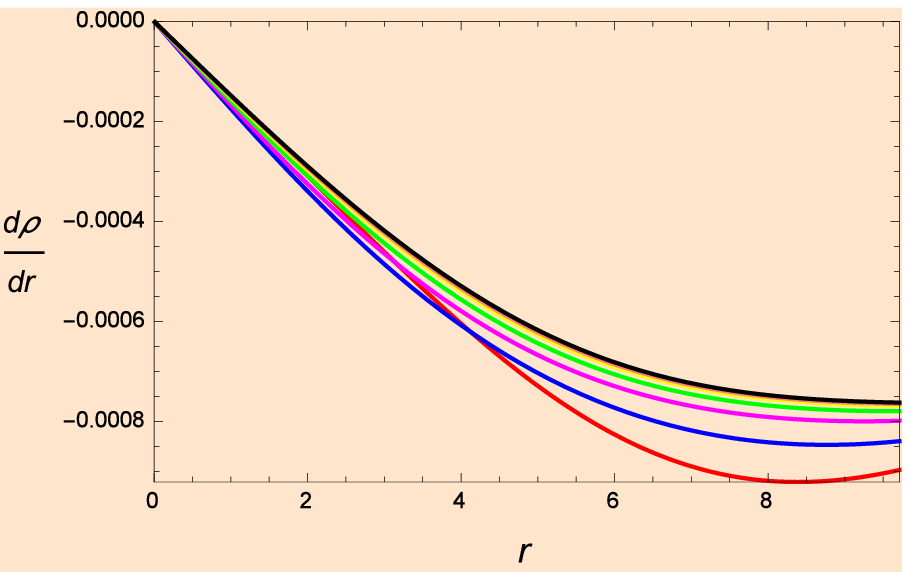, width=.32\linewidth,
height=2in}\epsfig{file=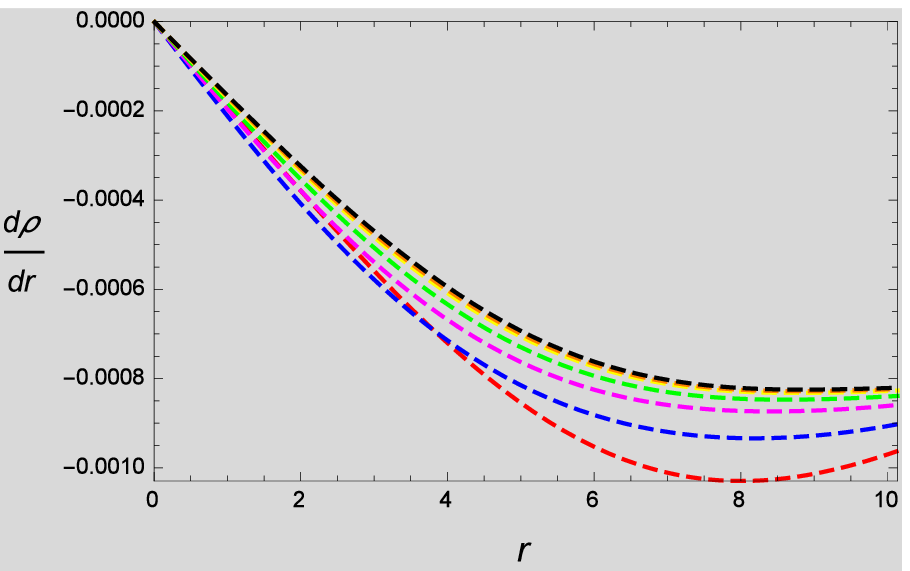, width=.32\linewidth,
height=2in}\epsfig{file=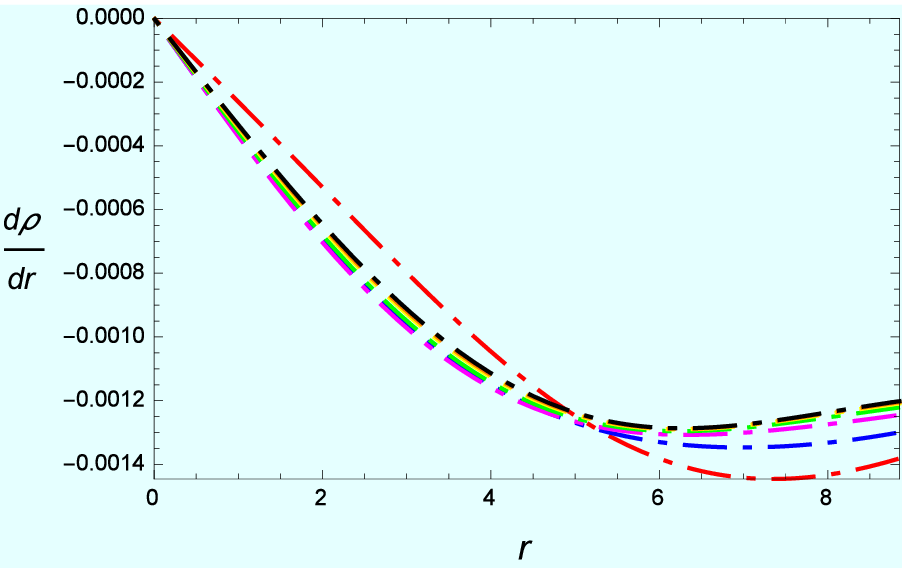, width=.32\linewidth,
height=2in}\caption{Visual representation of gradient of $\rho$ with $n=3(\textcolor{red}{\bigstar})$, $n=5(\textcolor{blue}{\bigstar})$, $n=10(\textcolor{magenta}{\bigstar})$, $n=20(\textcolor{green}{\bigstar})$, $n=50(\textcolor{yellow}{\bigstar})$, $n=100(\textcolor{orange}{\bigstar})$, and $n=500(\textcolor{black}{\bigstar})$}
\label{Fig.6}
\end{figure}

\begin{figure}[h]
\centering \epsfig{file=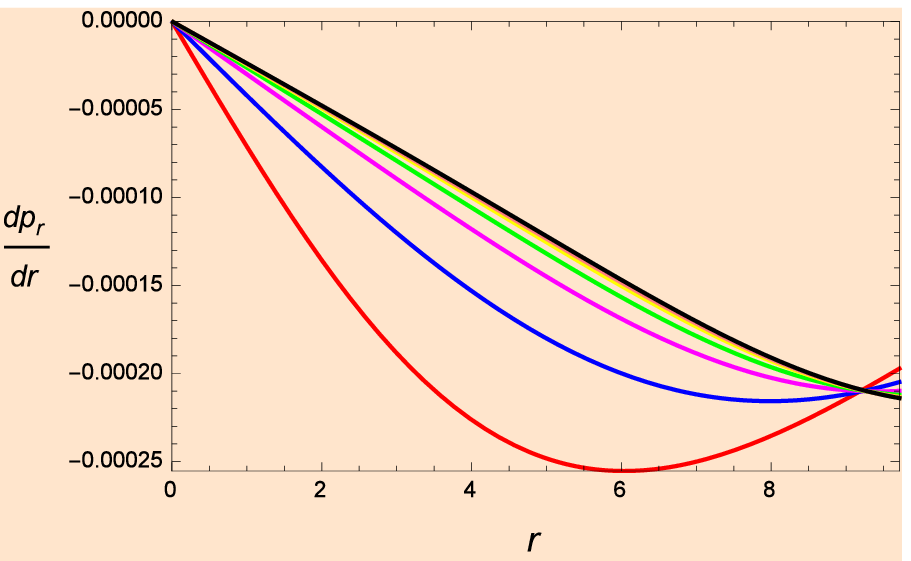, width=.32\linewidth,
height=2in}\epsfig{file=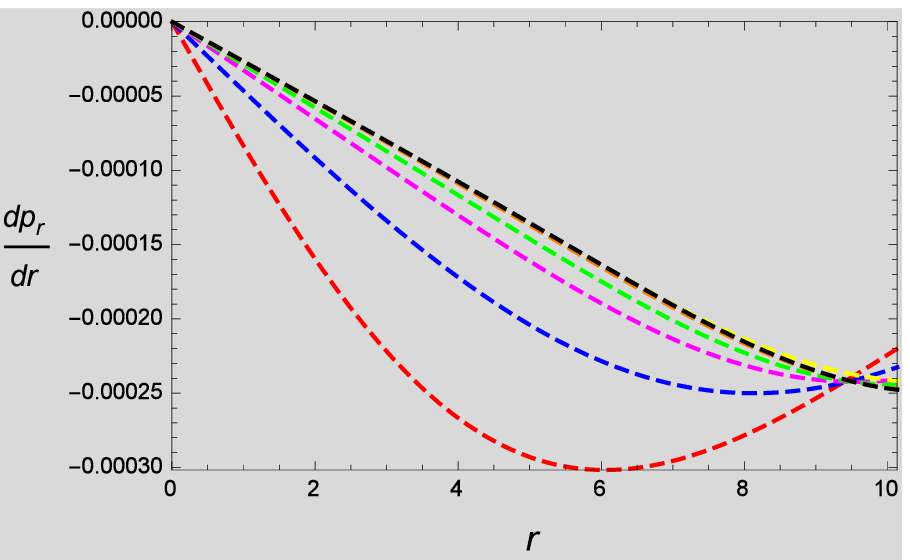, width=.32\linewidth,
height=2in}\epsfig{file=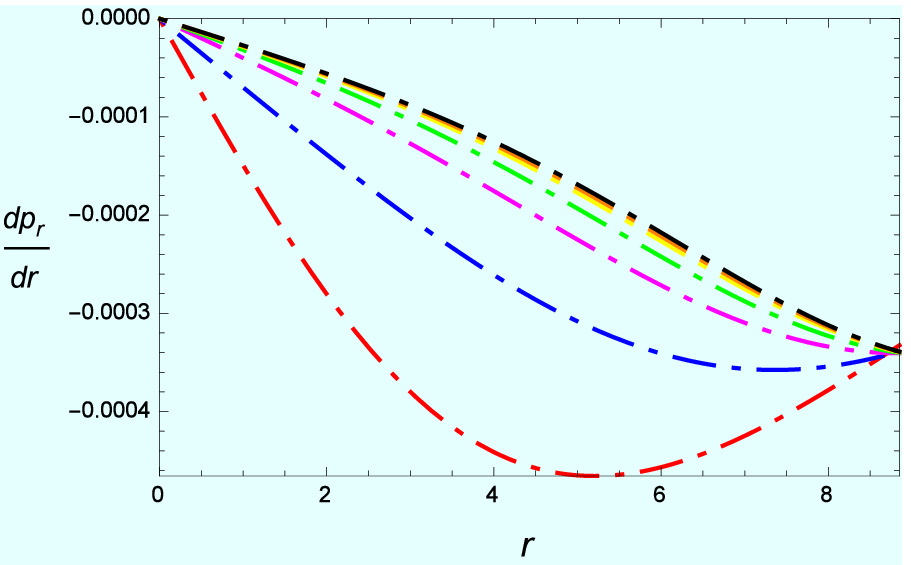, width=.32\linewidth,
height=2in}\caption{Visual representation of gradient of $p_r$ with $n=3(\textcolor{red}{\bigstar})$, $n=5(\textcolor{blue}{\bigstar})$, $n=10(\textcolor{magenta}{\bigstar})$, $n=20(\textcolor{green}{\bigstar})$, $n=50(\textcolor{yellow}{\bigstar})$, $n=100(\textcolor{orange}{\bigstar})$, and $n=500(\textcolor{black}{\bigstar})$}
\label{Fig.7}
\end{figure}

\begin{figure}[h]
\centering \epsfig{file=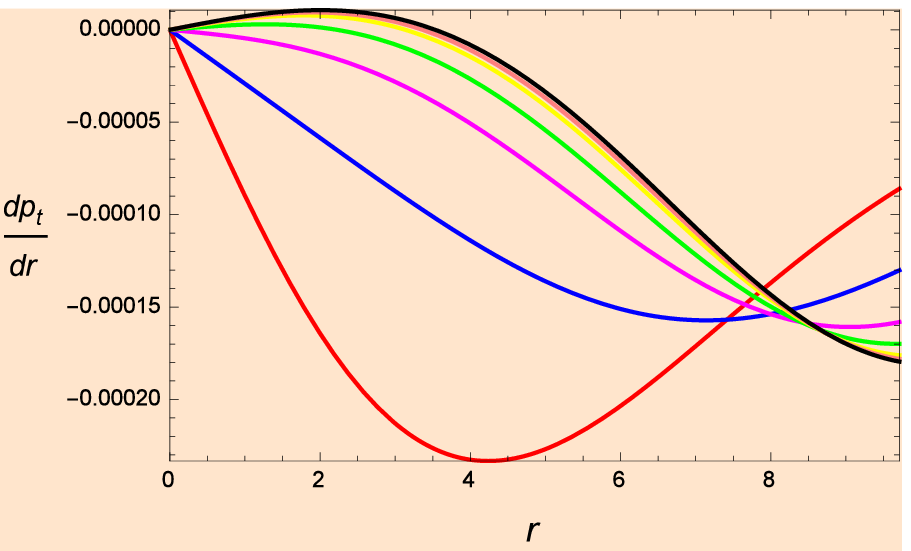, width=.32\linewidth,
height=2in}\epsfig{file=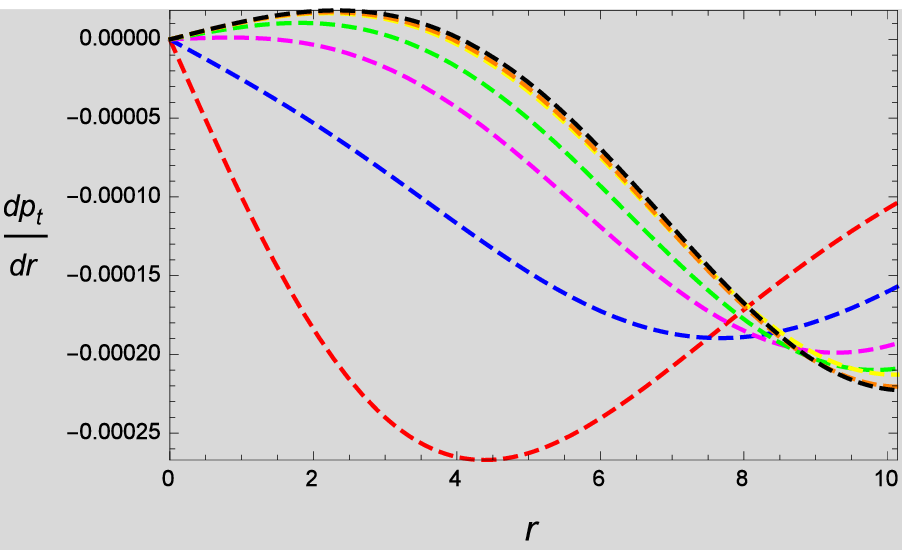, width=.32\linewidth,
height=2in}\epsfig{file=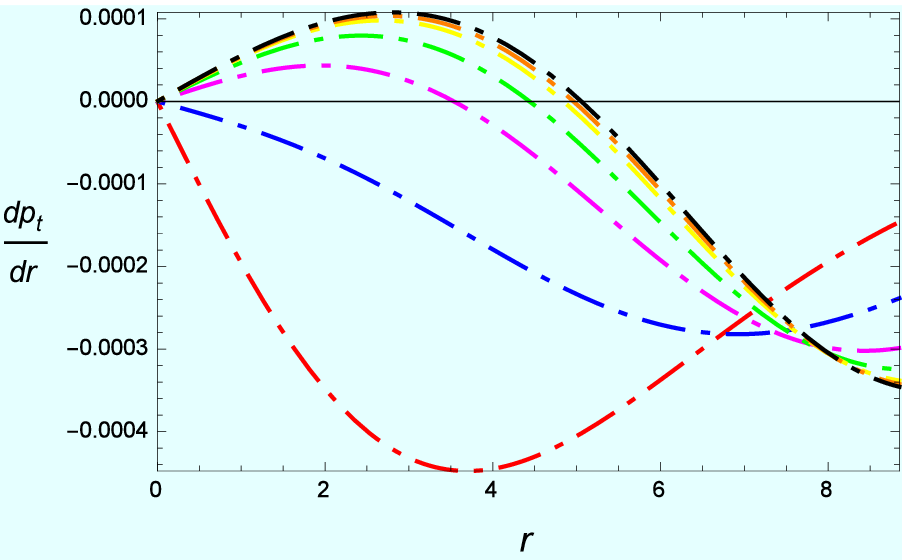, width=.32\linewidth,
height=2in}\caption{Visual representation of gradient of $p_t$ with $n=3(\textcolor{red}{\bigstar})$, $n=5(\textcolor{blue}{\bigstar})$, $n=10(\textcolor{magenta}{\bigstar})$, $n=20(\textcolor{green}{\bigstar})$, $n=50(\textcolor{yellow}{\bigstar})$, $n=100(\textcolor{orange}{\bigstar})$, and $n=500(\textcolor{black}{\bigstar})$}
\label{Fig.8}
\end{figure}

In order to model the interior geometry of the relativistic
stellar configuration under current circumstances, the role of
anisotropy is crucial for the compact sphere modeling and it is
represented as
\begin{equation}
\Delta=p_t-p_r.
\end{equation}
It depicts the information regarding the anisotropic nature of the
stellar configuration. If $p_t>p_r$  then the anisotropy is considered
to be non negative and is drawn outwards and depicted as $\Delta<0$.
Whereas if $p_r>p_t$ then the anisotropy turns out to be negative
and this shows that anisotropy is drawn inwards. From \ref{Fig.5} it is
observed that for our ongoing study anisotropy remains positive,
hence, directed outwards. The deviation of radial derivatives of
the energy density and pressure components , i.e., $\frac{d\rho}{dr}$, $\frac{dp_{r}}{dr}$ and $\frac{dp_{t}}{dr}$,
are shown in Figs. \ref{Fig.6}-\ref{Fig.8} such that\\
\begin{equation}\label{20}
\frac{d\rho}{dr}<0,~~~~~~~~~~~\frac{dp_{r}}{dr}<0,~~~~~~~~\frac{dp_{t}}{dr}<0.
\end{equation}
It can be observed from the second order derivatives that the
pressure components and energy density show the maximum
value at the core $r=0$, i. e.,
\begin{equation}\label{21}
\frac{d^2\rho}{dr^2}>0,~~~~~~~~~~~\frac{d^2p_{r}}{dr^2}>0,~~~~~~~~\frac{d^2p_{t}}{dr^2}>0.
\end{equation}


\begin{center}
\begin{table}
\caption{\label{tab1}{Predicted values of physical parameters at center and boundary.}}
\begin{tabular}{|c|c|c|c|c|c|c|c|c|}
    \hline
    & \multicolumn{6}{|c|}{LMC X-4} \\
\hline
n     &$e^{a(r=0)}$   &$e^{b(r=0)}$   &$\rho_{R}\;(g/cm^3)$      &$p_{r_{0}}=p_{t_{0}}\;(dyne/cm^{2})$      &$\rho_{0}\;(g/cm^3)$           &$p_{r_{0}}/\rho_{0}=p_{t_{0}}/\rho_{0}$\\
\hline
3   &0.431  &1.0   &$0.7546\times 10^{14}$   &$13.5220\times 10^{34}$        &$1.8378\times 10^{14}$        &$0.074$\\
5   &0.432  &1.0   &$0.8479\times 10^{14}$   &$14.1351\times 10^{34}$        &$1.7467\times 10^{14}$        &$0.081$\\
10  &0.433  &1.0   &$0.9735\times 10^{14}$   &$15.9780\times 10^{34}$        &$1.6629\times 10^{14}$        &$0.094$\\
20  &0.434  &1.0   &$1.0004\times 10^{14}$   &$16.1932\times 10^{34}$        &$1.5492\times 10^{14}$        &$0.110$\\
50  &0.440  &1.0   &$1.0875\times 10^{14}$   &$17.1260\times 10^{34}$        &$1.4207\times 10^{14}$        &$0.124$\\
100 &0.441  &1.0   &$1.1787\times 10^{14}$   &$18.1910\times 10^{34}$        &$1.3328\times 10^{14}$        &$0.135$\\
500 &0.442  &1.0   &$1.2433\times 10^{14}$   &$18.9280\times 10^{34}$        &$1.2558\times 10^{14}$        &$0.136$\\
\hline
\end{tabular}
\end{table}
\end{center}

\begin{center}
\begin{table}
\caption{\label{tab1}{Predicted values of physical parameters at center and boundary.}}
\begin{tabular}{|c|c|c|c|c|c|c|c|c|}
    \hline
    & \multicolumn{6}{|c|}{Cen X - 3 (mass =1.49$M/M_{\odot}$ \& radii= 10.136 km)} \\
\hline
n     &$e^{a(r=0)}$   &$e^{b(r=0)}$   &$\rho_{R}\;(g/cm^3)$      &$p_{r_{0}}=p_{t_{0}}\;(dyne/cm^{2})$      &$\rho_{0}\;(g/cm^3)$           &$p_{r_{0}}/\rho_{0}=p_{t_{0}}/\rho_{0}$\\
\hline
3   &0.376  &1.0   &$1.7236\times 10^{14}$   &$16.4320\times 10^{34}$         &$1.9578\times 10^{14}$        &$0.095$\\
5   &0.377  &1.0   &$1.8429\times 10^{14}$   &$17.1451\times 10^{34}$         &$1.8769\times 10^{14}$        &$0.103$\\
10  &0.378  &1.0   &$1.9325\times 10^{14}$   &$18.9380\times 10^{34}$         &$1.7629\times 10^{14}$        &$0.125$\\
20  &0.379  &1.0   &$2.0137\times 10^{14}$   &$19.1632\times 10^{34}$         &$1.5092\times 10^{14}$        &$0.140$\\
50  &0.380  &1.0   &$2.0945\times 10^{14}$   &$20.8460\times 10^{34}$         &$1.4007\times 10^{14}$        &$0.164$\\
100 &0.381  &1.0   &$2.3527\times 10^{14}$   &$21.1220\times 10^{34}$         &$1.4428\times 10^{14}$        &$0.169$\\
500 &0.382  &1.0   &$2.5723\times 10^{14}$   &$21.9180\times 10^{34}$         &$1.4958\times 10^{14}$        &$0.172$\\
\hline
\end{tabular}
\end{table}
\end{center}

\begin{center}
\begin{table}
\caption{\label{tab1}{Predicted values of physical parameters at center and boundary.}}
\begin{tabular}{|c|c|c|c|c|c|c|c|c|}
\hline
 & \multicolumn{6}{|c|}{EXO 1785-248 (Mass =1.30$M/M_{\odot}$ \& Radii=8.849 km)} \\
\hline
n     &$e^{a(r=0)}$   &$e^{b(r=0)}$   &$\rho_{R}\;(g/cm^3)$      &$p_{r_{0}}=p_{t_{0}}\;(dyne/cm^{2})$     &$\rho_{0}\;(g/cm^3)$           &$p_{r_{0}}/\rho_{0}=p_{t_{0}}/\rho_{0}$\\
\hline
3   &0.376    &1.0     &$0.8876\times 10^{14}$    &$15.4634\times 10^{34}$         &$1.9684\times 10^{14}$    &$0.082$\\
5   &0.381    &1.0     &$0.9574\times 10^{14}$    &$16.7635\times 10^{34}$         &$1.8073\times 10^{14}$    &$0.099$\\
10  &0.384    &1.0     &$1.0165\times 10^{14}$    &$17.3650\times 10^{34}$         &$1.7871\times 10^{14}$    &$0.114$\\
20  &0.385    &1.0     &$1.0969\times 10^{14}$    &$18.7532\times 10^{34}$         &$1.6063\times 10^{14}$    &$0.131$\\
50  &0.386    &1.0     &$1.1607\times 10^{14}$    &$19.8591\times 10^{34}$         &$1.4534\times 10^{14}$    &$0.146$\\
100 &0.387    &1.0     &$1.2485\times 10^{14}$    &$20.1813\times 10^{34}$         &$1.3115\times 10^{14}$    &$0.153$\\
500 &0.388    &1.0     &$1.4524\times 10^{14}$    &$20.9994\times 10^{34}$         &$1.3089\times 10^{14}$    &$0.156$\\
\hline
\end{tabular}
\end{table}
\end{center}

\subsection{Energy Conditions}
\begin{figure}[h]
\centering \epsfig{file=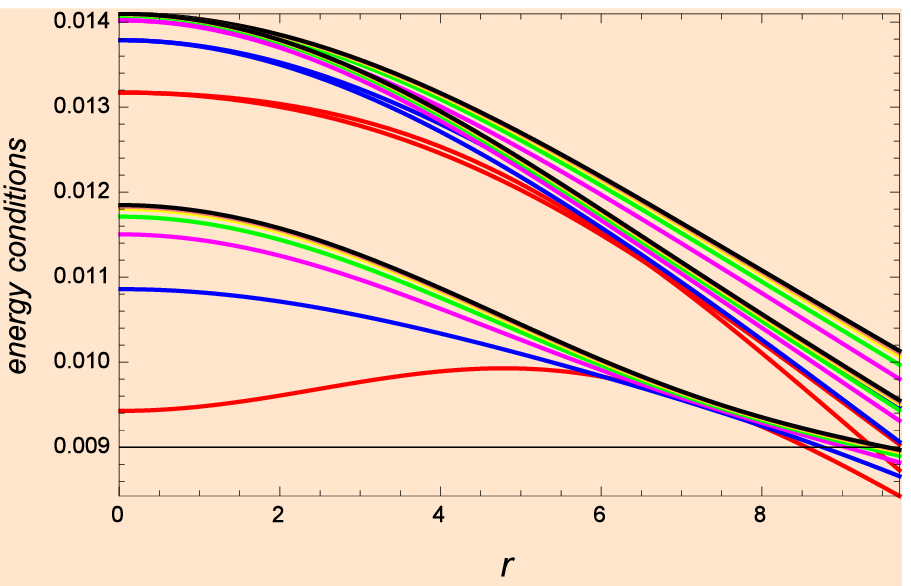, width=.32\linewidth,
height=2in}\epsfig{file=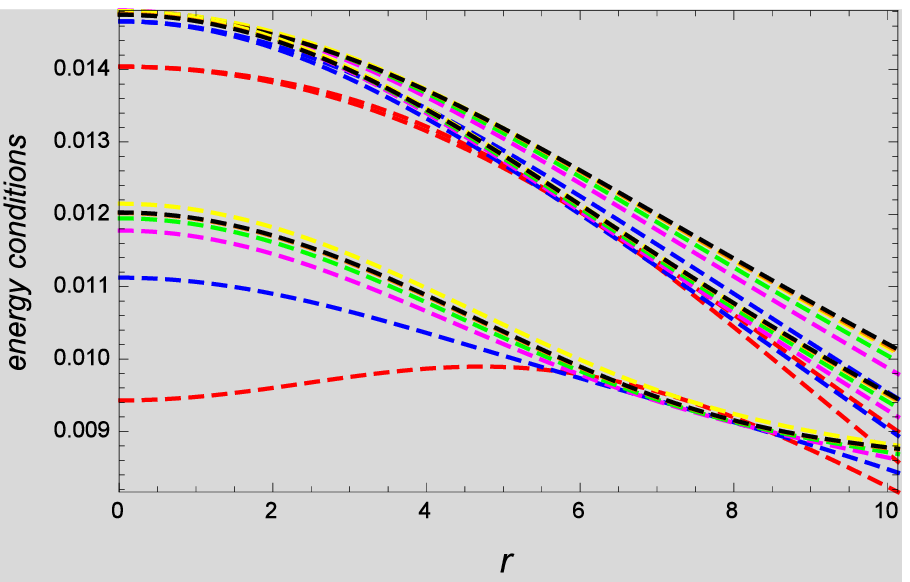, width=.32\linewidth,
height=2in}\epsfig{file=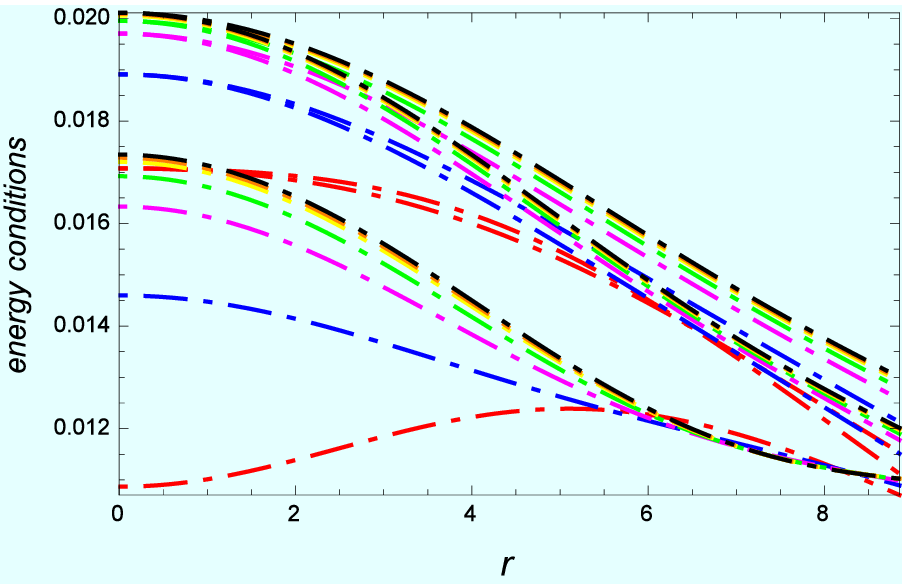, width=.32\linewidth,
height=2in}\caption{Evolution of energy bounds with $n=3(\textcolor{red}{\bigstar})$, $n=5(\textcolor{blue}{\bigstar})$, $n=10(\textcolor{magenta}{\bigstar})$, $n=20(\textcolor{green}{\bigstar})$, $n=50(\textcolor{yellow}{\bigstar})$, $n=100(\textcolor{orange}{\bigstar})$, and $n=500(\textcolor{black}{\bigstar})$}
\label{Fig.9}
\end{figure}

Energy conditions appear to be quite helpful in analyzing
the realistic distribution of matter. These attributes play a decisive role to classify the exotic and normal mater distribution
within the stellar model. The energy conditions have been crucially important in debating the issues related to cosmology and
astrophysics. The energy condition are classified as
\begin{eqnarray}
\nonumber
&&NEC:\rho>0,\;\;\;\rho+p_r\geq0,~~~\rho+p_t\geq0 \nonumber,~~~DEC:\rho\geq|p_r|,~~~\rho\geq|p_t|\nonumber,
\\&&WEC:~~~\rho-p_r\geq0,~~~\rho-p_t\geq0 \nonumber,
\\&&SEC:~~~\rho-p_r\geq0,~~~\rho-p_t\geq0,~~~\rho-p_r-2p_t\geq0.
\end{eqnarray}
Here, $NEC$ stands for null energy condition, $SEC$ for strong
energy condition , $DEC$ for dominant energy condition and
$WEC$ for week energy condition. It can be seen from Fig. \ref{Fig.9}
that all the energy bounds exhibit decreasing behavior with the
increase in radii of the compact stellar sphere.
\subsection{Analysis of Stability}

\begin{figure}[h]
\centering \epsfig{file=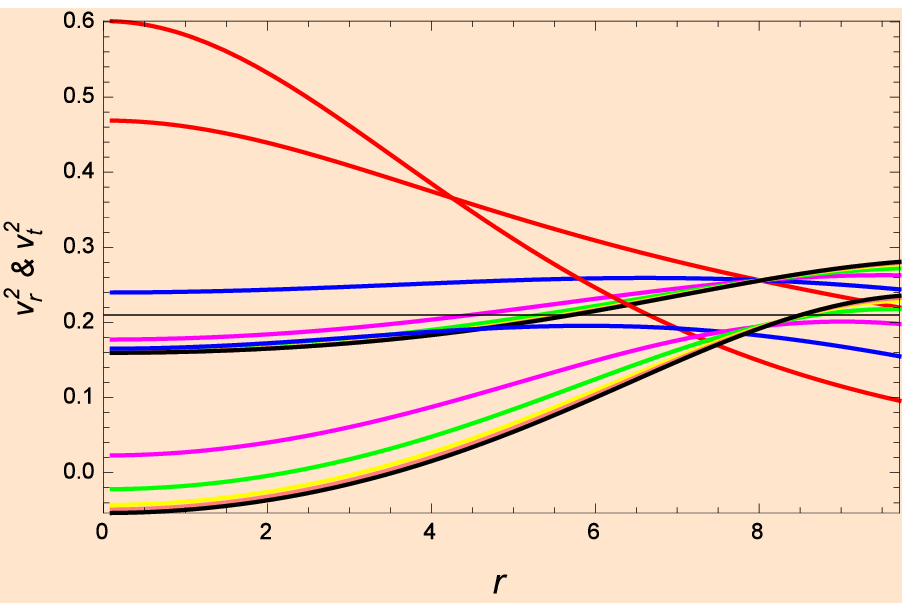, width=.32\linewidth,
height=2in}\epsfig{file=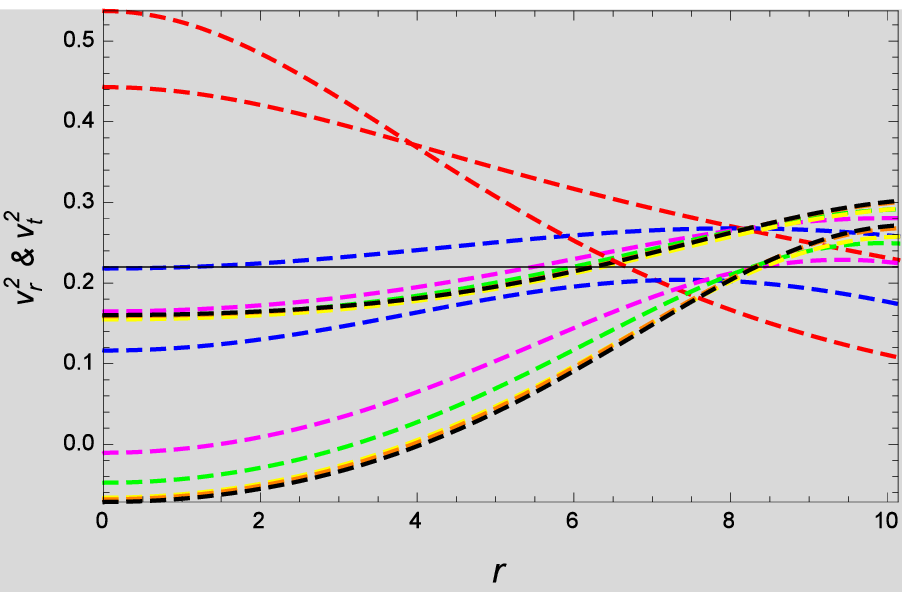, width=.32\linewidth,
height=2in}\epsfig{file=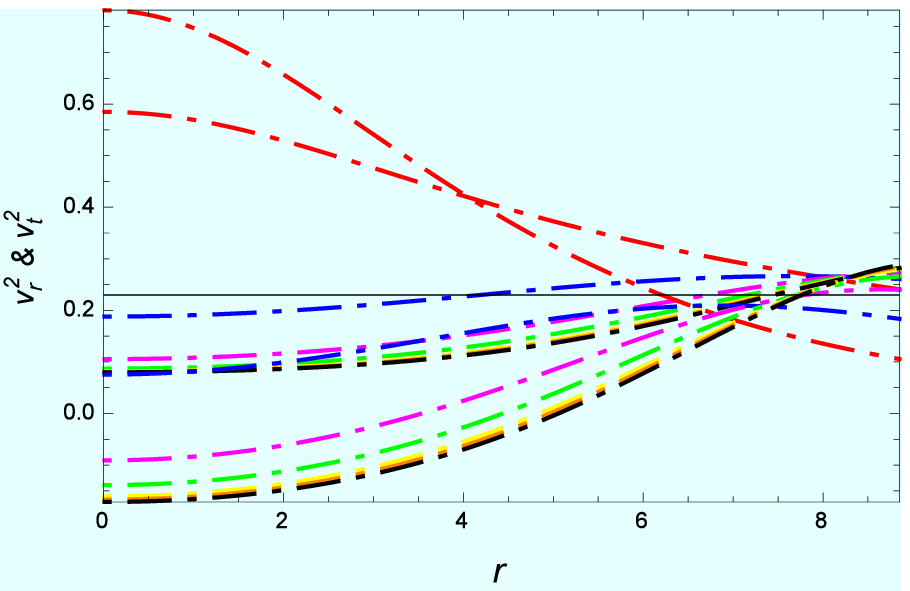, width=.32\linewidth,
height=2in}\caption{Visual representation of $v^2_{r}$ and $v^2_{t}$ with $n=3(\textcolor{red}{\bigstar})$, $n=5(\textcolor{blue}{\bigstar})$, $n=10(\textcolor{magenta}{\bigstar})$, $n=20(\textcolor{green}{\bigstar})$, $n=50(\textcolor{yellow}{\bigstar})$, $n=100(\textcolor{orange}{\bigstar})$, and $n=500(\textcolor{black}{\bigstar})$}
\label{Fig.10}
\end{figure}

\begin{figure}[h]
\centering \epsfig{file=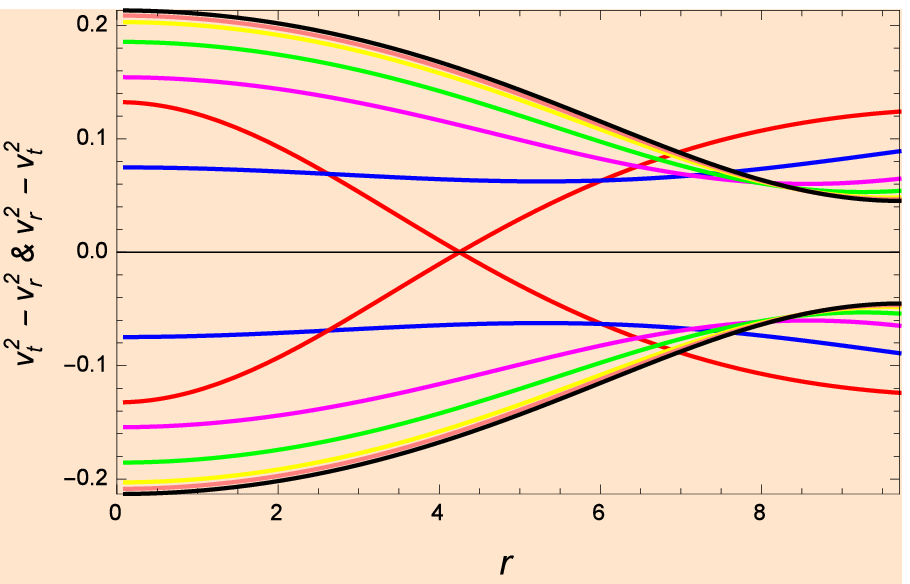, width=.32\linewidth,
height=2in}\epsfig{file=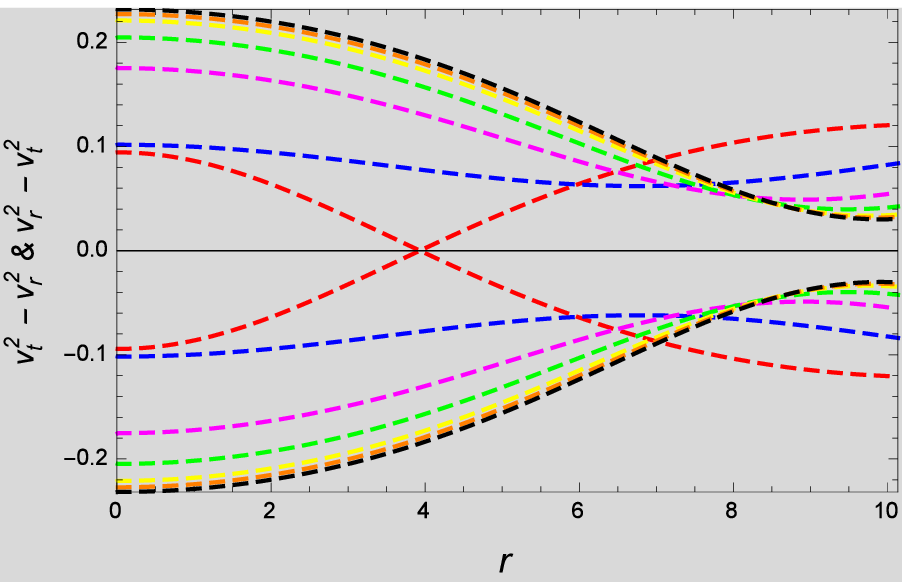, width=.32\linewidth,
height=2in}\epsfig{file=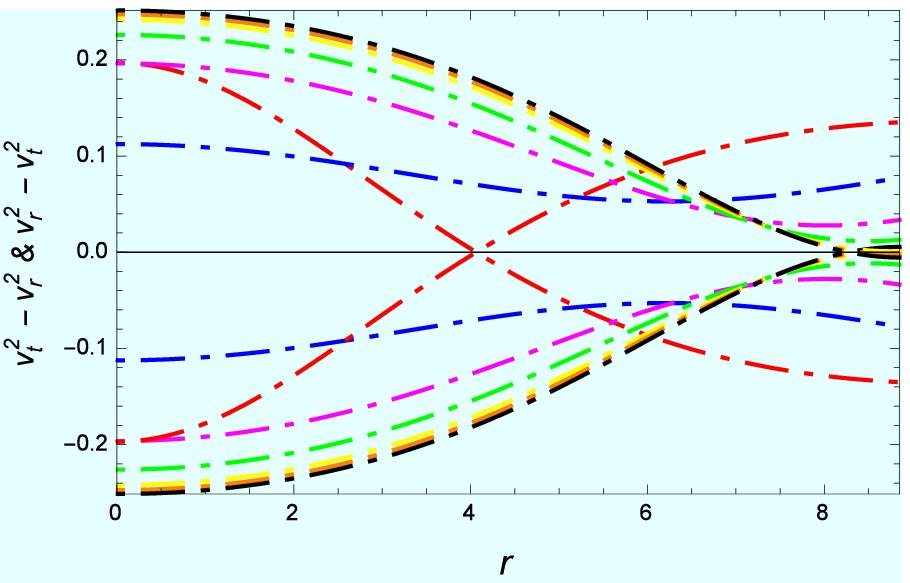, width=.32\linewidth,
height=2in}\caption{Visual representation of $|v^2_{t}-v^2_{r}|$ and $|v^2_{r}-v^2_{t}|$ with $n=3(\textcolor{red}{\bigstar})$, $n=5(\textcolor{blue}{\bigstar})$, $n=10(\textcolor{magenta}{\bigstar})$, $n=20(\textcolor{green}{\bigstar})$, $n=50(\textcolor{yellow}{\bigstar})$, $n=100(\textcolor{orange}{\bigstar})$, and $n=500(\textcolor{black}{\bigstar})$}
\label{Fig.11}
\end{figure}

The stability of the stellar configuration plays a decisive role in
analyzing the consistency of the acquired model. Many analytical
discussions have been done in order to find the stability of the
matter configuration but Herrera's cracking conception emerged
to be very effective \cite{Herrera}. The radial and tangential speed of sound are
defined as
\begin{equation}\label{28}
v^2_{sr}=\frac{dp_{r}}{d\rho},~~~~~\text{and}~~~~~~ v^2_{st}=\frac{dp_{t}}{d\rho}.
\end{equation}
For the conservation of causality condition, components of the speed of sound must be with the bounds of the interval $[0,1]$  i.e $0\leq v^2_{r}~and~ v^2_{t}\leq1$. It can be observed from the Figs. \ref{Fig.10}-\ref{Fig.11} that the condition i.e $0\leq v^2_{r}~and~ v^2_{t}\leq1$ is satisfied by both velocity components. Apart from that Abreu condition i.e $-1\leq |v^2_{t}-v^2_{r}|\leq0$ has also been satisfied and can be observed from the Fig. \ref{Fig.11}. The validity of both of the aspects affirm the viability and the effectiveness of our model. Further, it can also be observed that inverse Abreu condition i.e $0\leq |v^2_{t}-v^2_{r}|\leq1$ is also satisfied.
\subsection{Equilibrium Analysis for Modified $f(\mathcal{R,T})$ Gravity }
\begin{figure}[h]
\centering \epsfig{file=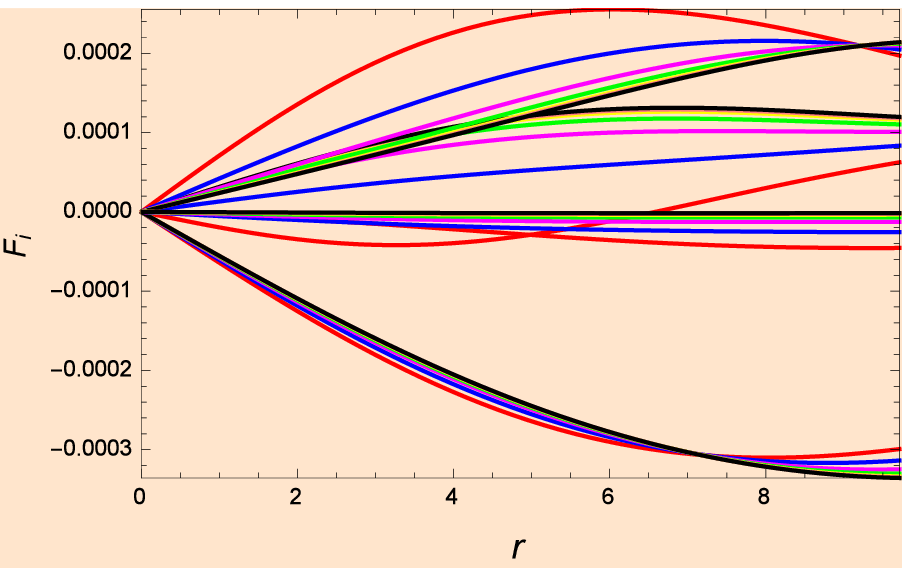, width=.32\linewidth,
height=2in}\epsfig{file=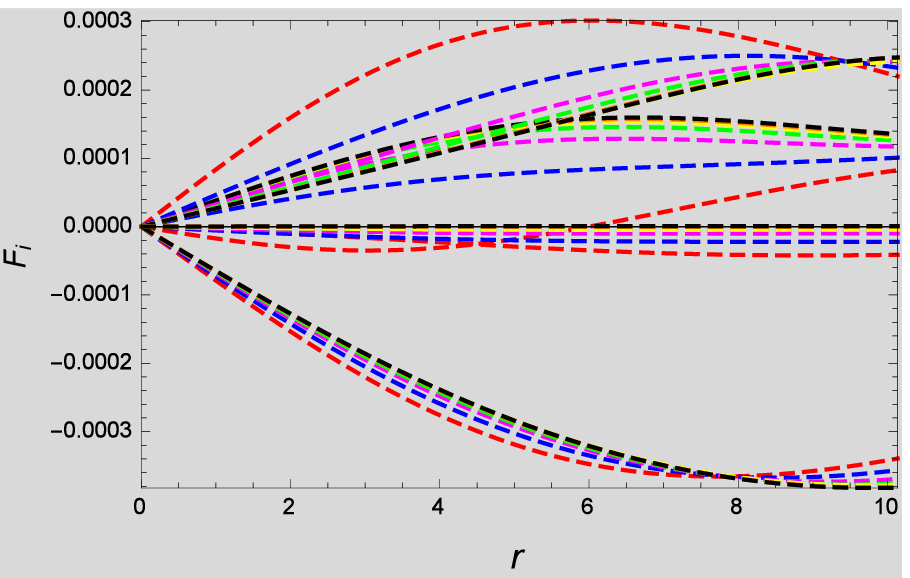, width=.32\linewidth,
height=2in}\epsfig{file=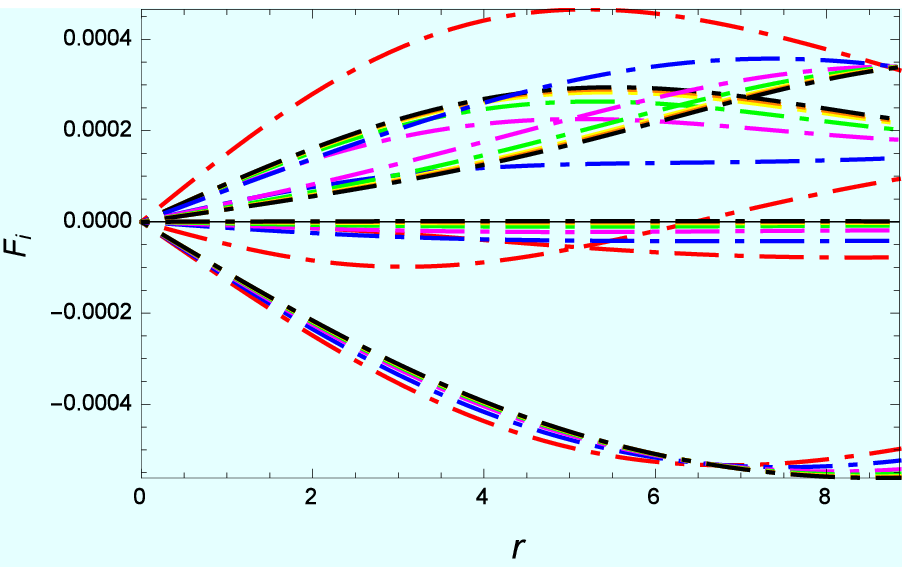, width=.32\linewidth,
height=2in}\caption{Visual representation of $\mathcal{F}_g$, $\mathcal{F}_h$, $\mathcal{F}_a$, and $\mathcal{F}_e$ with $n=3(\textcolor{red}{\bigstar})$, $n=5(\textcolor{blue}{\bigstar})$, $n=10(\textcolor{magenta}{\bigstar})$, $n=20(\textcolor{green}{\bigstar})$, $n=50(\textcolor{yellow}{\bigstar})$, $n=100(\textcolor{orange}{\bigstar})$, and $n=500(\textcolor{black}{\bigstar})$}
\label{Fig.12}
\end{figure}

In this section, we will analyze the equilibrium condition by
considering the stability of the acquired solution of the three
different stellar configuration. For the purpose, we make use of
the TOV equation \cite{35}-\cite{40}
\begin{equation}
\frac{2\Delta}{r}-\frac{dp_{r}}{dr}-\frac{a'}{2}(\rho +p_r)+\frac{\lambda  \left(-\frac{d p_{r}}{d r}-2 \frac{d p_{t}}{d r}+3 \frac{d \rho }{d r}\right)}{3 (2 \lambda +1)}=0.
\end{equation}
The above equation characterizes the necessary and sufficient
condition for the hydrostatic-equilibrium. It comprises of four
different forces
\begin{equation}
\mathcal{F}_g= \frac{a'}{2}(\rho +p_r),~~~\mathcal{F}_h= \frac{dp_{r}}{dr},~~~\mathcal{F}_a= \frac{2\Delta}{r},\;\;\mathcal{F}_{e}=\frac{\lambda  \left(-\frac{d p_{r}}{d r}-2 \frac{d p_{t}}{d r}+3 \frac{d \rho }{d r}\right)}{3 (2 \lambda +1)}.
\end{equation}
\begin{itemize}
\item $\mathcal{F}_a$ represents the anisotropy force.
\item $\mathcal{F}_h$ represents the hydrostatic force.
\item $\mathcal{F}_g$ represents the gravitational force.
\item $\mathcal{F}_e$ represents the extra force.
\end{itemize}
Consequently, the $TOV$ equation can also be written as $\mathcal{F}_g +\mathcal{F}_h+\mathcal{F}_a +\mathcal{F}_e=0$. From the attained graph as shown in Fig. \ref{Fig.12}, it is deduced that all of the forces sum up to neutralize the total effect, and this confirms the existence of the stable stellar structures.
\subsection{Evolution of Adiabatic Index}

\begin{figure}[h]
\centering \epsfig{file=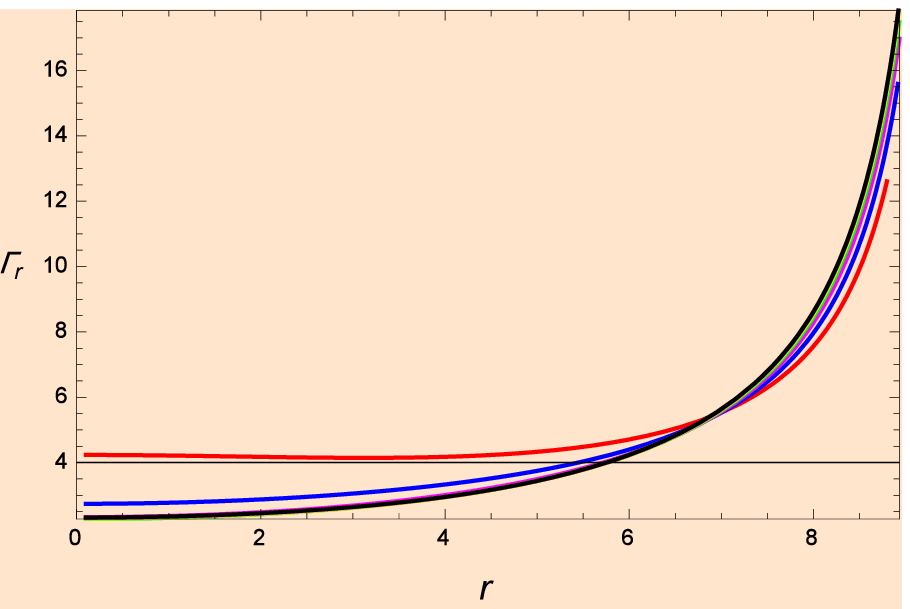, width=.32\linewidth,
height=2in}\epsfig{file=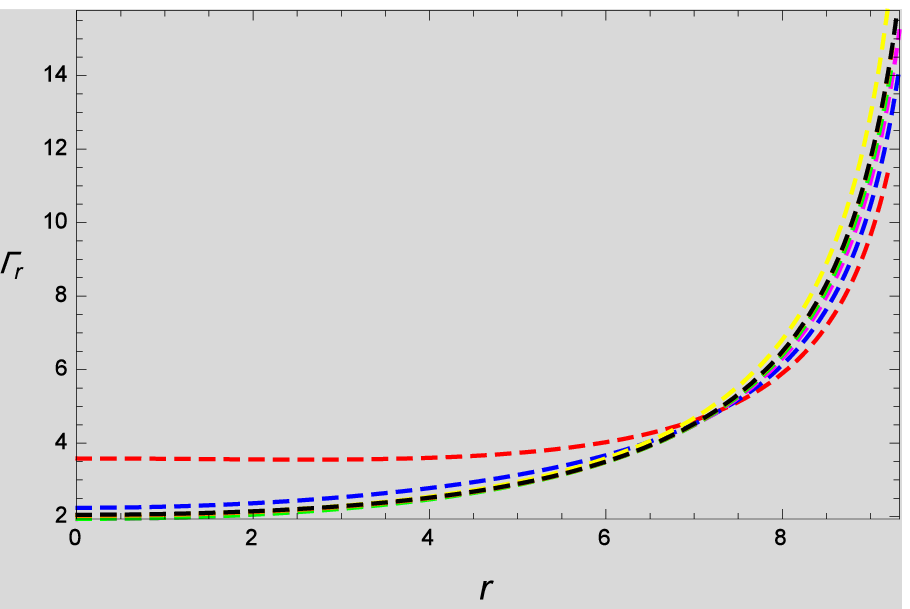, width=.32\linewidth,
height=2in}\epsfig{file=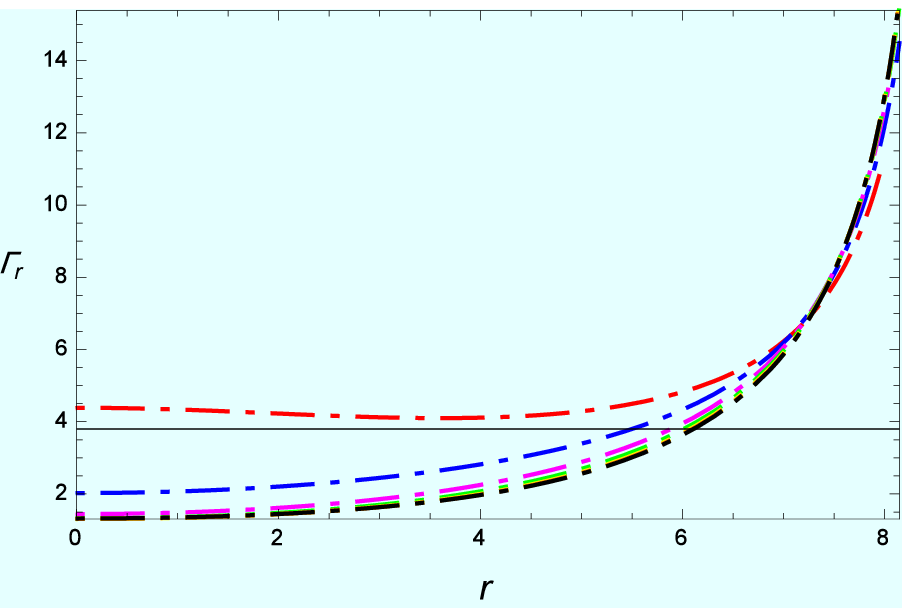, width=.32\linewidth,
height=2in}\caption{Evolution of $\Gamma_r$ with $n=3(\textcolor{red}{\bigstar})$, $n=5(\textcolor{blue}{\bigstar})$, $n=10(\textcolor{magenta}{\bigstar})$, $n=20(\textcolor{green}{\bigstar})$, $n=50(\textcolor{yellow}{\bigstar})$, $n=100(\textcolor{orange}{\bigstar})$, and $n=500(\textcolor{black}{\bigstar})$}
\label{Fig.13}
\end{figure}

For the energy density, EoS stiffness can be better described
by the adiabatic index. The stability of relativistic as well as the
non-relativistic stellar structures can be explained through the
adiabatic index. The concept of the dynamical stability via the
radial adiabatic index was presented by Chandrasekhar \cite{41}. This
was further utilized by many authors \cite{42}-\cite{47}. For the system to
be dynamically stable, the adiabatic index must go beyond $4/3$.
Adiabatic index corresponding to the radial stress is given as
\begin{equation}\label{26}
\Gamma_r=\frac{\rho+p_{r}}{p_r}(\frac{dp_{r}}{d\rho})= \frac{\rho+p_{r}}{p_r}v^2_{r}.
\end{equation}
One of the quite fascinating fact of the above equation is that the stability of the Newtonian matter configuration is achieved when $\Gamma_r > 4/3$. While if  $\Gamma_r = 4/3$ then a neutral equilibrium is achieved whereas if $\Gamma_r < 4/3$ then an unstable matter configuration consisting of anisotropy is achieved. Anisotropic matter profile via adiabatic index can be elaborated as
\begin{equation}\label{26}
\Gamma_r=\frac{4}{3} +(\frac{\rho_i p_{ri}}{2\mid p'_{ri}\mid}r+\frac{4}{3}\frac{p_{ti}-p_{ri}}{\mid p'_{ri}\mid r} ).
\end{equation}
From the Fig. \ref{Fig.13}, the graphical behavior of $\Gamma_r$ with respect to increasing radii can be observed. It is noted that $\Gamma_r$ shows the monotonically increasing conduct for all the stellar spheres and $\Gamma_r$ is always greater than $4/3$. Hence, $\Gamma_r$ is consistent for the stability of our model in $f(\mathcal{R}, \mathcal{T})$ gravity.
\subsection{Equation of state}

The evolution of the emergence of compact stars can be determined by $EoS$ of matter. Moreover, $EoS$ has a strong impact
on the conditions of nucleosynthesis. Therefore, $EoS$ is a vital
tool in many astrophysical simulations. The $EoS$ is considered
to be a ratio of the pressure terms $p_r$ and $p_t$ with density. The
components of $EoS$ for the study of stellar configuration are i.e.,
$\omega_r$ and $\omega_t$ and are mathematically connected as
\begin{equation}
 \omega_r=\frac{p_r}{\rho}, ~~~~~~  \omega_t=\frac{p_t}{\rho}.
\end{equation}
\begin{figure}[h]
\centering \epsfig{file=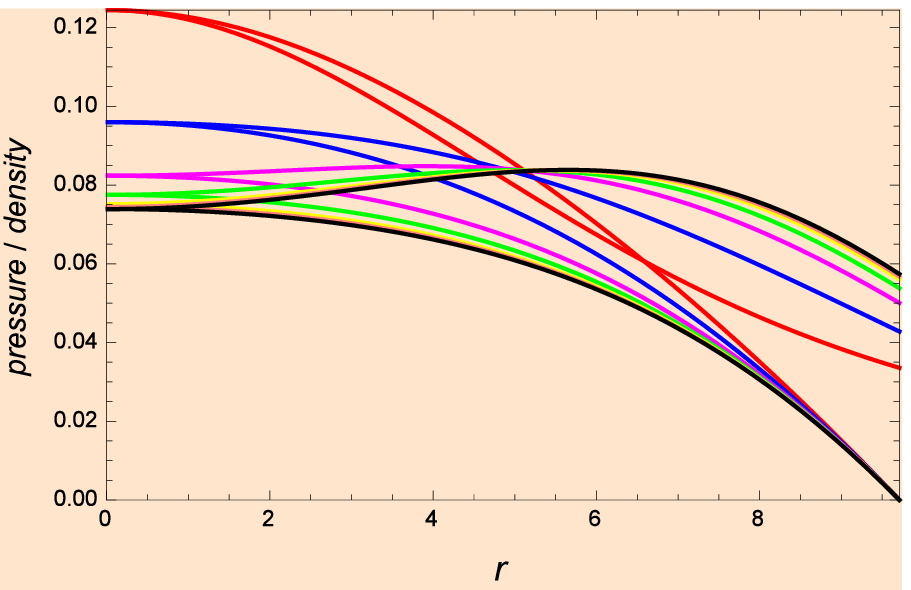, width=.32\linewidth,
height=2in}\epsfig{file=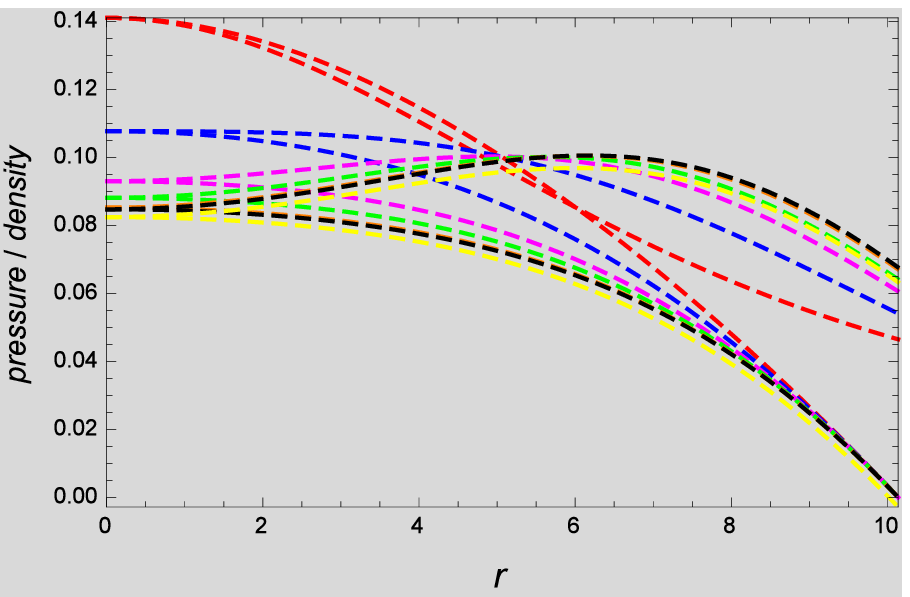, width=.32\linewidth,
height=2in}\epsfig{file=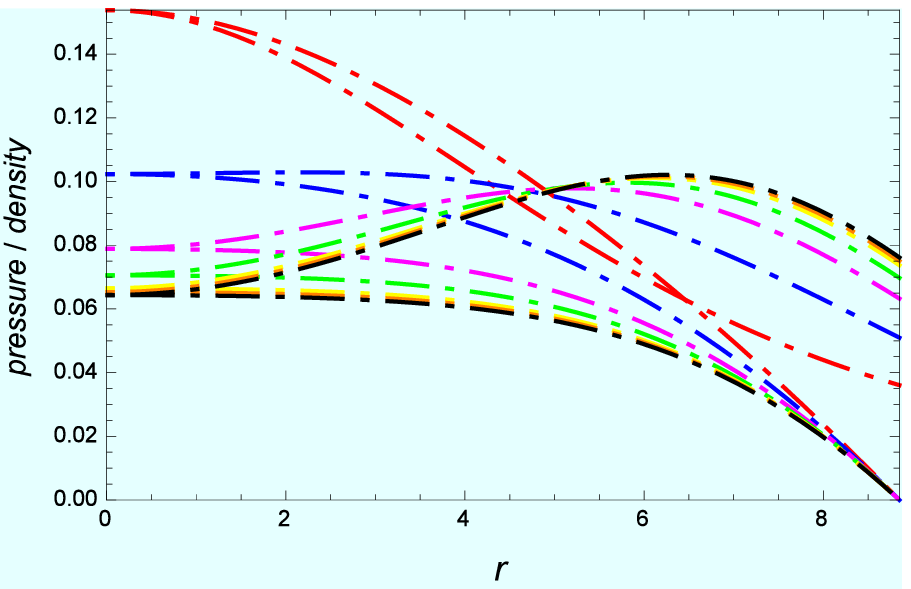, width=.32\linewidth,
height=2in}\caption{Evolution of energy $EoS$ with $n=3(\textcolor{red}{\bigstar})$, $n=5(\textcolor{blue}{\bigstar})$, $n=10(\textcolor{magenta}{\bigstar})$, $n=20(\textcolor{green}{\bigstar})$, $n=50(\textcolor{yellow}{\bigstar})$, $n=100(\textcolor{orange}{\bigstar})$, and $n=500(\textcolor{black}{\bigstar})$}
\label{Fig.14}
\end{figure}
From Fig. \ref{Fig.14}, it can be observed that with the increase in radii, the components of the $EoS$ shows monotonically decreasing behavior and are always less than 1. Moreover, the positive nature is observed for both of  the components of $EoS$ i.e. $\omega_r$ and $\omega_t$ with in the matter configuration. The accomplishment of the condition i.e. $0 \leq\omega_r$ and $\omega_t < 1$ unveils that the our obtained solutions are valid and legitimate.

\begin{figure}[b]
\centering \epsfig{file=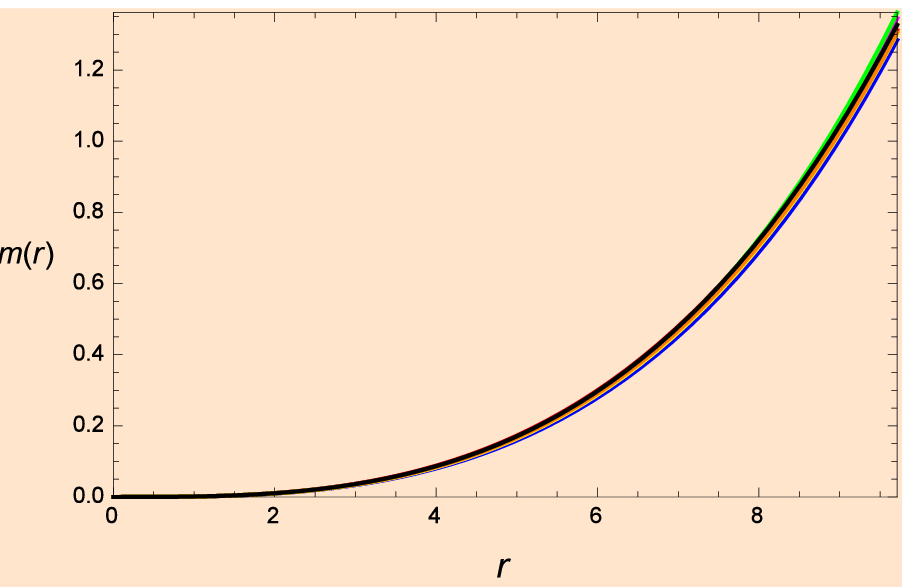, width=.32\linewidth,
height=2in}\epsfig{file=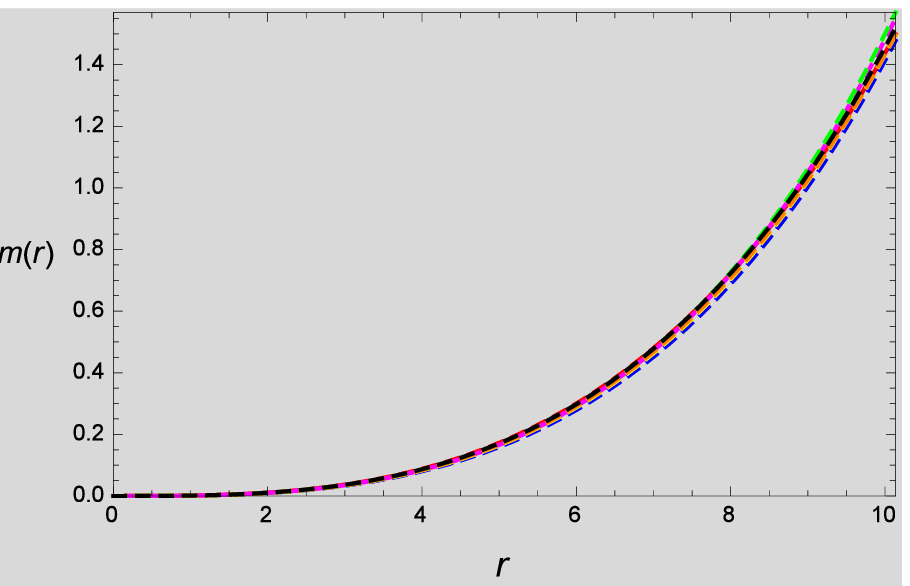, width=.32\linewidth,
height=2in}\epsfig{file=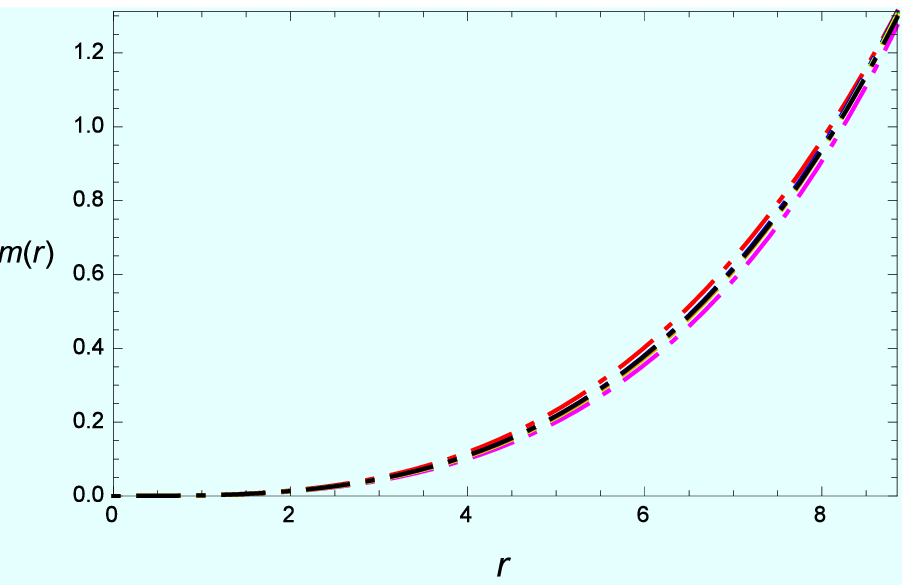, width=.32\linewidth,
height=2in}\caption{Evolution of $m(r)$ with $n=3(\textcolor{red}{\bigstar})$, $n=5(\textcolor{blue}{\bigstar})$, $n=10(\textcolor{magenta}{\bigstar})$, $n=20(\textcolor{green}{\bigstar})$, $n=50(\textcolor{yellow}{\bigstar})$, $n=100(\textcolor{orange}{\bigstar})$, and $n=500(\textcolor{black}{\bigstar})$}
\label{Fig.15}
\end{figure}
\begin{figure}[h]
\centering \epsfig{file=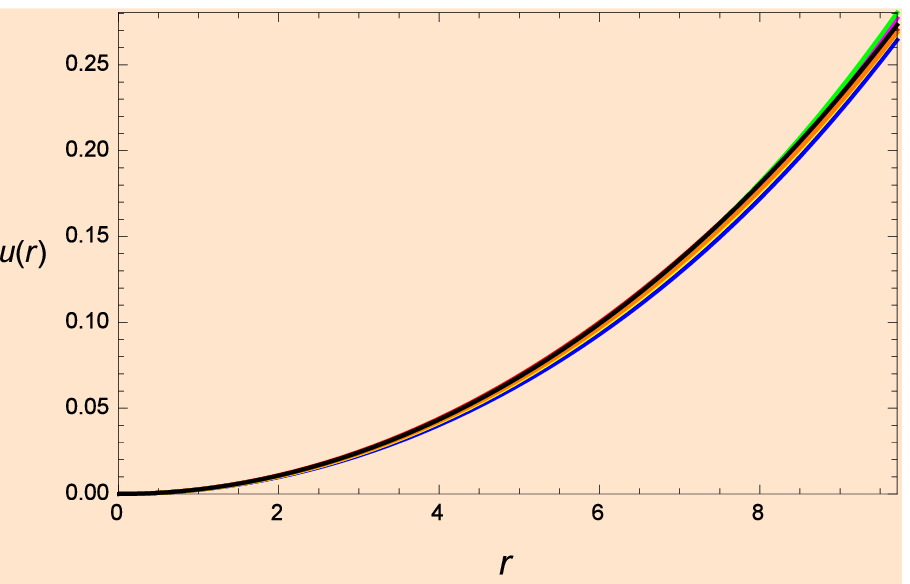, width=.32\linewidth,
height=2in}\epsfig{file=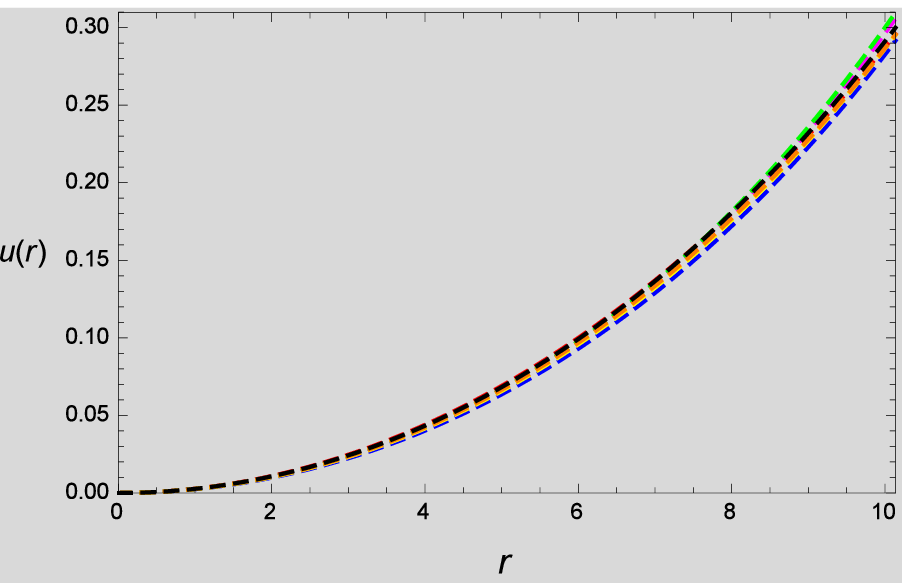, width=.32\linewidth,
height=2in}\epsfig{file=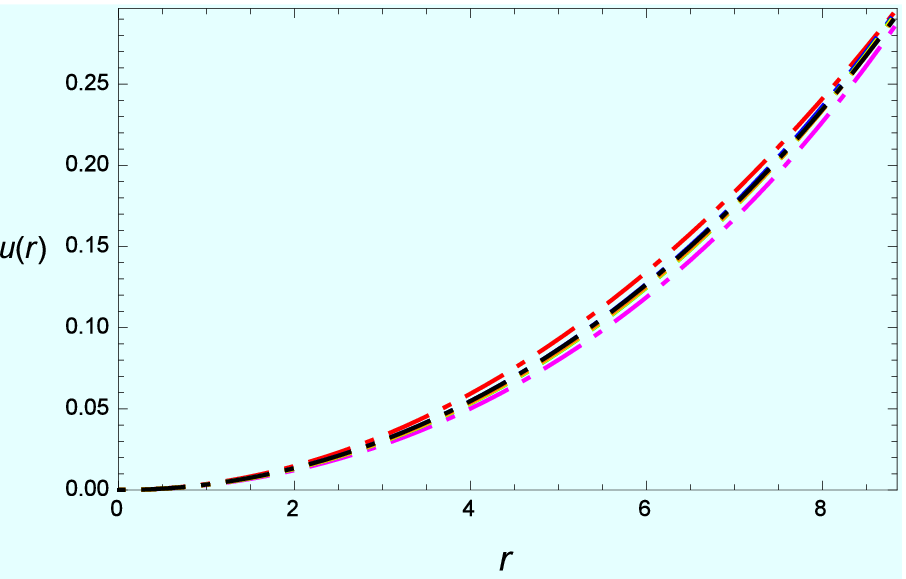, width=.32\linewidth,
height=2in}\caption{Evolution of $\mu(r)$ with $n=3(\textcolor{red}{\bigstar})$, $n=5(\textcolor{blue}{\bigstar})$, $n=10(\textcolor{magenta}{\bigstar})$, $n=20(\textcolor{green}{\bigstar})$, $n=50(\textcolor{yellow}{\bigstar})$, $n=100(\textcolor{orange}{\bigstar})$, and $n=500(\textcolor{black}{\bigstar})$}
\label{Fig.16}
\end{figure}
\begin{figure}[h]
\centering \epsfig{file=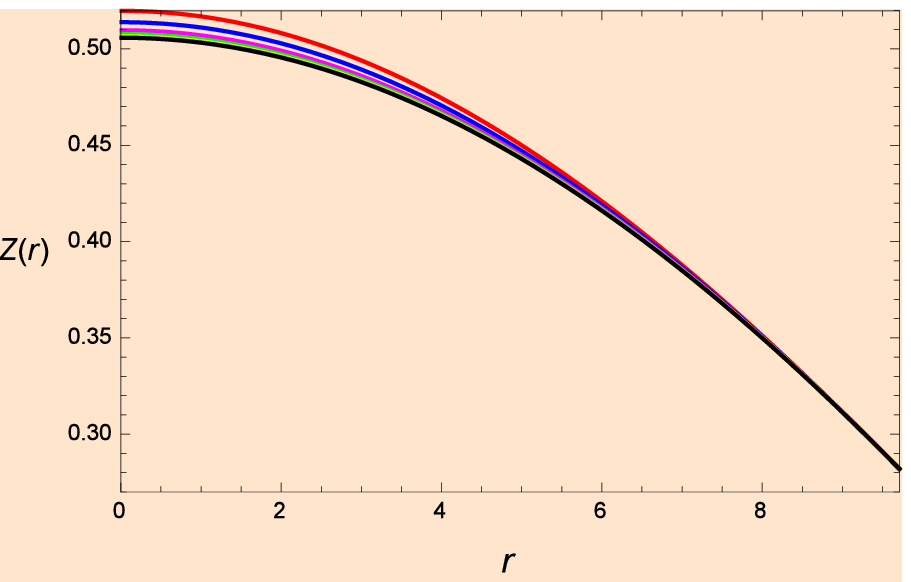, width=.32\linewidth,
height=2in}\epsfig{file=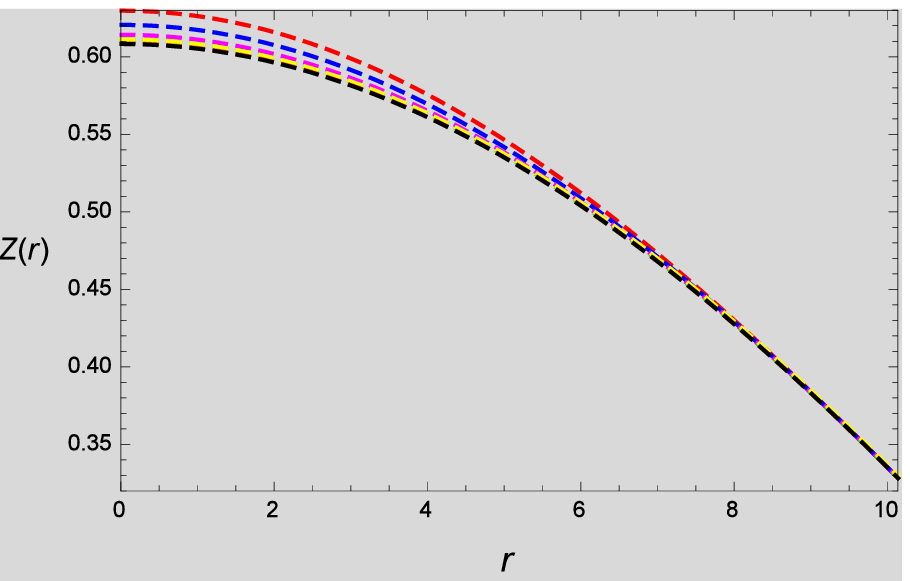, width=.32\linewidth,
height=2in}\epsfig{file=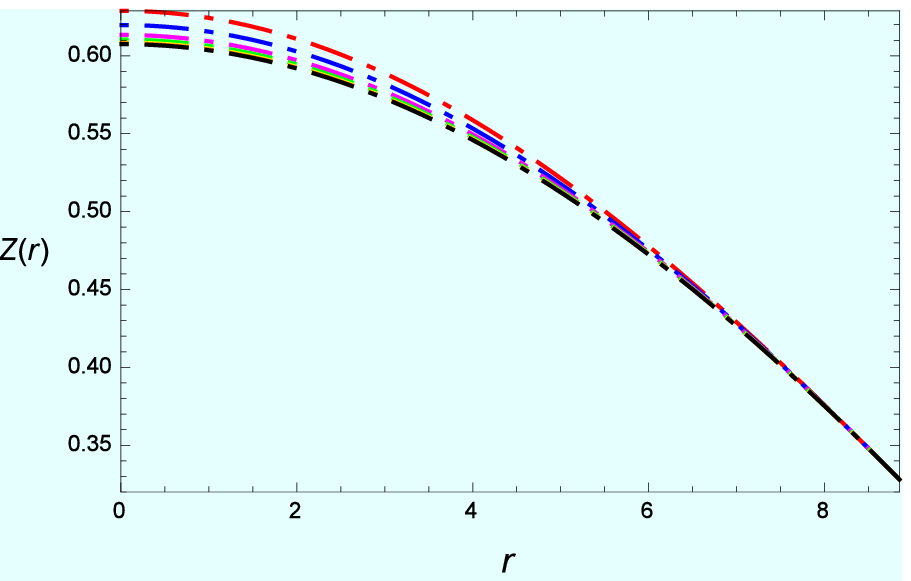, width=.32\linewidth,
height=2in}\caption{Evolution of $z_{s}$ with $n=3(\textcolor{red}{\bigstar})$, $n=5(\textcolor{blue}{\bigstar})$, $n=10(\textcolor{magenta}{\bigstar})$, $n=20(\textcolor{green}{\bigstar})$, $n=50(\textcolor{yellow}{\bigstar})$, $n=100(\textcolor{orange}{\bigstar})$, and $n=500(\textcolor{black}{\bigstar})$}
\label{Fig.17}
\end{figure}

\subsection{Compactness factor and Surface Redshift}
For the existence of any matter configuration mass function,
compactness factor along with surface redshift function are contemplated to be an essential constituent. The fundamental relation for the mass function is given as
\begin{equation}
m(r)= \int_{0} ^{R} 4\pi \rho r^2 dr.
\end{equation}
The generalize compactness factor i.e. $\mu(r)$ is represented as
\begin{equation}
\mu(r)=\frac{m}{r} =\frac{1}{r}\int_{0} ^{R} 4\pi \rho r^2 dr.
\end{equation}
The strong intermolecular interaction forces with in the stellar
matter configuration and its corresponding $EoS$ can be characterized by the term surface redshift i.e., $Z_s$. The generalized relation
is
\begin{equation}
Z_s=\frac{1}{(\psi _1 \left(\psi _2 r^2+1\right){}^n)^{1/2}}-1.
\end{equation}
Figs. \ref{Fig.15}-\ref{Fig.17}
represent the evolution of redshift function along
with compactness factor corresponding to the increasing radii. It
can be observed that the surface redshift is always less than 5
i.e., $5$ i.e. $Z_s\leq5$ and the compactness factor remains less than $0.30$ i.e.,  $\mu(r)\leq0.30$. All of the mentioned functions are positive
throughout the configuration. Hence, our models are stable.

\section{Conclusion}

The manifesto of the current study is to identify the realistic
and stable configuration for the stellar sphere in the modified
$f(\mathcal{R}, \mathcal{T})$ gravitational theory. For the analysis of the stellar matter
configuration, a viable model $f(\mathcal{R}, \mathcal{T})=\mathcal{R} + \beta \mathcal{R}^2 + \lambda \mathcal{T}$ is considered along with spherically symmetric space–time. In order to achieve the current objective, observational data of three compact
spheres i.e. LMC X-4, Cen X-3 and EXO 1785–248 inheriting an
anisotropic matter distribution has been utilized. This realistic
range for masses of stars under this study is 1.29 to 1.5 solar
mass. In this context, the considered models LMC X-4 (mass 1.29 $M/M_{\odot}$), Cen X-3 (mass 1.49 $M/M_{\odot}$) and EXO 1785-248 (mass 1.29 $M/M_{\odot}$) are with in suitable range under this study. Moreover, the radii are also with in prescribed range 8.849 km to
10.136 km. In general, the study is valid for other models of stars
and those within the given ranges of mass and radius under this
study. Embedding class 1 condition is used to find the potential
i.e. $g_{rr}$ by considering the primary potential as i.e $g_{tt}=e^{a(r)}=\chi_{1} \left(r^{2} \chi_{2}+1\right)^{K}$. To find the unknown constraints, Schwarzschild’s exterior solution has been utilized. All the obtained results can be summarized as:
\begin{itemize}
\item  {\it Metric potentials}:  The presence of singularities within the stellar configuration is an essential topic, worthy of debate. The stability of the stellar matter configuration depends on it. Therefore, the aspects of metric potential play a decisive role. The graphical behavior of Fig. \ref{Fig.1} depicts that the fundamental condition  i.e. $ e^{b(r=0)}=1$  and $e^{a(r=0)}\neq 0$ has been encompassed by the gravitational potential. Increasing attribute of the potentials is also observed throughout the configuration. Therefore, the potentials are free from any singularity and so is our model.

\item  {\it Energy density and stress constraints}: From the Figs. \ref{Fig.2}-\ref{Fig.4} the evolution of density and components of stresses i.e., $p_r$ and $p_t$ can be observed. The density along with stress components show decreasing behavior and are non-negative throughout the configuration. The peak value is accomplished at the core while decreasing evolution is observed with the increase in radii and it tends to 0 towards the boundary.

\item  {\it Anisotropy and gradients}: Fig. \ref{Fig.5} depicts anisotropy behavior for our current configuration. It is noted that $p_t\neq p_r$  and $p_{t}>p_{r}$, therefore, $\bigtriangleup>0$ so the anisotropy is positive and is directed outwards. From Fig. \ref{Fig.6}-\ref{Fig.8} gradient of density and stress components are reviewed and it is noticed that all the gradients are negative and exhibit decreasing behavior i.e.,
 $\frac{d\rho}{dr}<0$,$\frac{dp_{r}}{dr}<0$, $\frac{dp_{t}}{dr}<0$. Since the non-positive behavior of all gradients along with their vanishing attribute at $r=0$ is observed, therefore, our stellar configuration is stable.

\item  {\it Energy bounds}: From the Figs. \ref{Fig.2}-\ref{Fig.4} and Fig. \ref{Fig.9} the characteristic behavior of functions $\rho$, $p_r$, $p_t$, $\rho-p_r$, $\rho-p_t$, $\rho-p_r-2p_t$ are depicted. Decreasing characteristic behavior is observed with the increase in radii which further fulfills the bounds of energy such as $(NEC)$, $(SEC)$, $(DEC)$ and $(WEC)$. Hence the matter profile is realistic and viable.

\item  {\it Causality analysis}   The behavior of the constraints of sound speed is depicted in Figs. \ref{Fig.10} and Fig. \ref{Fig.11}. The decreasing attribute is observed for both the components of the speed and it is shown that they are always within the limits i.e., $0\leq v^2_{r}\leq1$ and $0\leq v^2_{t}\leq1$. The fulfillment of Aberu condition is also observed i.e., $-1\leq |v^2_{t}-v^2_{r}|\leq0$. For the current model $v^2_{t}>v^2_{r}$. Therefore, fulfillment of all the conditions confirms the viability of the stellar sphere.

\item  {\it Equilibrium and $EoS$ analysis}: The balancing nature of all the forces, i.e., $\mathcal{F}_a$, $\mathcal{F}_h$ and $\mathcal{F}_g$ is depicted in Fig. \ref{Fig.12}. As all these forces add up to 0 and balance the effect of each other, therefore, the equilibrium condition is satisfied. From Fig. \ref{Fig.14}, the attributes of the parameters of $EoS$ are observed. It is concluded that constraints i.e., $\omega_r$ and $\omega_t$ of $EoS$ are positive in the interior of stellar profile and are in the stability bounds of $0 \leq\omega_r$ and $\omega_t < 1$.

\item {\it Adiabatic index stability analysis}: The behavior of adiabatic index $\Gamma_r$ can be seen in Fig. \ref{Fig.13} showing that $\Gamma_r > 4/3 $. It also depicts the positive and decreasing nature, justifying the effectiveness of our system in the framework of $f(\mathcal{R}, \mathcal{T})$ theory.

\item  {\it Redshift, mass function and compactness factor}:  Fig. \ref{Fig.15}-\ref{Fig.17} exhibit the compactness factor, mass function along with gravitational redshift. It can be seen that both $\mu(r)\leq0.30$ and $m(r)$ show increasing behavior. Apart from this,  $Z_s$ shows the decreasing attribute and  $Z_s\leq5$ which is in alliance with the stability of the configuration.
\end{itemize}
It is worth mentioning here that our obtained solutions in current study represent more dense stellar structures as compared to past related works on compact objects in $f(\mathcal{R}, \mathcal{T})$ gravity \cite{23,25,26,48,
49}.
\section*{Appendix (\textbf{I})}

\begin{eqnarray*}
\varUpsilon _1&&=r^2 \psi _2+1,\;\;\;\;\varUpsilon _2=\frac{48 \beta  (3 \lambda +2) r^4 \psi _3^3 \psi _2^2 \left((n-1) r^2 \psi _2+1\right){}^3}{\left(\varUpsilon _1^{3-n}+r^2 \psi _2 \psi _3 \varUpsilon _1\right){}^3},\;\;\;\;\varUpsilon _3=\varUpsilon _1 \left((n-1) r^2 \psi _2-n+4\right),\\
\varUpsilon _4&&=\varUpsilon _1 \left((n-1) (2 n-3) r^2 \psi _2+5 n-8\right),\;\varUpsilon _5=r^2 \psi _2 \psi _3 \varUpsilon _1^n+\varUpsilon _1^2,\;\;\varUpsilon _6=r^2 \psi _2 \left((n-1) (2 n-3) r^2 \psi _2+9 n-15\right)+6,\\
\varUpsilon _7&&=r^2 \psi _2 \left((n-1) (n+1) (2 n-3) r^4 \psi _2^2+3 (n (n+4)-9) r^2 \psi _2+6 n-3\right)+3,\\
\varUpsilon _8&&=r^2 \psi _2 \left((n-1) r^2 \psi _2-3 n+9\right)+3,\;\;\varUpsilon _9=\frac{2 \beta  r^2 \psi _3^2 \psi _2 \left((n-1) r^2 \psi _2+1\right){}^2 \left(48 \lambda -\frac{22 (3 \lambda +2) n r^2 \psi _2 \left(-\varUpsilon _1\right)}{\varUpsilon _1^2}+24\right)}{\left(\varUpsilon _1^{3-n}+r^2 \psi _2 \psi _3 \varUpsilon _1\right){}^2},\\
\varUpsilon _{10}&&=\lambda +\frac{8 \beta  (3 \lambda +2) \psi _3 \psi _2 \varUpsilon _1^{n-2} \left(\varUpsilon _1^2 \varUpsilon _4-r^2 \psi _3 \psi _2 \varUpsilon _3 \varUpsilon _1^n\right)}{\varUpsilon _5^2}+\lambda  r^2 \psi _2 \psi _3 \varUpsilon _1^{n-2},\\
\varUpsilon _{11}&&=\frac{4 \beta  n^3 r^4 \psi _2^2 \left(\frac{2 r \psi _3 \psi _2 (6 \lambda  r+r) \left((n-1) r^2 \psi _2+1\right)}{\varUpsilon _1^{3-n}+r^2 \psi _2 \psi _3 \varUpsilon _1}-4 (3 \lambda +1)\right)}{\varUpsilon _1^3},\\
\varUpsilon _{12}&&=4 \psi _3 \left((n-1) r^2 \psi _2+1\right) \left(\left((2 \lambda +1) r^2-6 \beta  \lambda \right) \left(r^2 \psi _2 \psi _3 \varUpsilon _1^{n-2}+1\right)+\beta  \left(-6 \lambda +\frac{2}{\varUpsilon _1^3}\times (3 \lambda +2) r^2 \psi _2 \left(-8 n \right.\right.\right.\\&&\times\left.\left.\left.\left(r^4 \psi _2^2-1\right)+6 n r^2 \psi _2 \left(r^2 \psi _2-3\right)-\frac{7 \psi _3 \varUpsilon _1^{n+1} \left(\varUpsilon _1^2 \varUpsilon _4-r^2 \psi _3 \psi _2 \varUpsilon _3 \varUpsilon _1^n\right)}{\varUpsilon _5^2}\right)-8\right)\right),
\end{eqnarray*}
\begin{eqnarray*}
\varUpsilon _{13}&&=8 \beta  (3 \lambda +2)+2 \lambda  r^2 \left(r^2 \psi _2 \psi _3 \varUpsilon _1^{n-2}+1\right)+2 \beta  r \psi _2 \left(\frac{(8-24 \lambda ) r \psi _3 \left((n-1) r^2 \psi _2+1\right)}{\varUpsilon _1^{3-n}+r^2 \psi _2 \psi _3 \varUpsilon _1}\right.\\&&-\left.\frac{2 (9 \lambda +11) r^3 \psi _3^2 \psi _2 \left((n-1) r^2 \psi _2+1\right){}^2}{\left(\varUpsilon _1^{3-n}+r^2 \psi _2 \psi _3 \varUpsilon _1\right){}^2}+\frac{2 (3 \lambda +4) r \psi _3 \varUpsilon _{15} \varUpsilon _1^{n-2}}{\varUpsilon _5^2}-\frac{4 n \left(-\varUpsilon _1\right) (6 \lambda  r+r)}{\varUpsilon _1^2}\right),\\
\varUpsilon _{14}&&=r^2 \psi _2 \psi _3 \varUpsilon _1^{n-2}+1,\;\;\;\varUpsilon _{15}=(n-1) r^2 \psi _2+1,\;\;\;\varUpsilon _{16}=\varUpsilon _1^2 \varUpsilon _4-r^2 \psi _3 \psi _2 \varUpsilon _3 \varUpsilon _1^n,\\
\varUpsilon _{17}&&=-\frac{24 (\lambda +1) \psi _3 \varUpsilon _{16} \varUpsilon _1^{n-2}}{\varUpsilon _5^2}+\frac{4 (3 \lambda -2) n \left(-\varUpsilon _1\right)}{\varUpsilon _1^2}-\frac{12 (3 \lambda +2) r^4 \psi _3^3 \psi _2^2 \varUpsilon _{15}^3}{\left(\varUpsilon _1^{3-n}+r^2 \psi _2 \psi _3 \varUpsilon _1\right){}^3},\\
\varUpsilon _{18}&&=\frac{4 (3 \lambda +4) n r^2 \psi _2 \left(r^2 \psi _2-3\right)}{\varUpsilon _1^3}+\frac{4 (3 \lambda +2) r^2 \psi _3 \psi _2 \varUpsilon _1^{n-3} \left(r^2 \psi _2 \psi _3^2 \left(\varUpsilon _1 \varUpsilon _8\right) \varUpsilon _1^{2 n}-\psi _3 \varUpsilon _7 \varUpsilon _1^{n+2}+(n-2) \varUpsilon _6 \varUpsilon _1^4\right)}{\varUpsilon _5^3},\\
\varUpsilon _{19}&&=\frac{8 (9 \lambda +8) r^2 \psi _2 \psi _3^2 \varUpsilon _{15}^2}{\left(\varUpsilon _1^{3-n}+r^2 \psi _2 \psi _3 \varUpsilon _1\right){}^2}+\psi _3 \varUpsilon _{15} \left(\frac{6 (5 \lambda +6) n r^2 \psi _2 \left(-\varUpsilon _1\right)}{\varUpsilon _1^2}+(3 \lambda +2) \left(\frac{14 r^2 \psi _3 \psi _2 \varUpsilon _{16} \varUpsilon _1^{n-2}}{\varUpsilon _5^2}+4\right)\right),
\end{eqnarray*}
\begin{eqnarray*}
\varUpsilon _{20}&&=\frac{8 \beta  (3 \lambda +1) n^4 r^7 \psi _2^4}{\varUpsilon _1^4}+\frac{8 \beta  n^3 r^5 \psi _2^3 \left(4 \lambda +\frac{2 r \psi _2 \psi _3 \varUpsilon _{15} (r-2 \lambda  r)}{\varUpsilon _1^{3-n}+r^2 \psi _2 \psi _3 \varUpsilon _1}\right)}{\varUpsilon _1^3}+\lambda  \left(\frac{2 n r^3 \psi _2 \left(-\varUpsilon _1\right)}{\varUpsilon _1^2}+\frac{24 \beta  r \psi _2 \psi _3 \varUpsilon _{15}}{\varUpsilon _1^{3-n}+r^2 \psi _2 \psi _3 \varUpsilon _1}\right),\\
\varUpsilon _{21}&&=-\frac{2 (5 \lambda +1) n^2 r^2 \psi _2 \left(-\varUpsilon _1\right){}^2}{\varUpsilon _1^4}-\frac{4 (7 \lambda +4) \psi _3 \varUpsilon _{16} \varUpsilon _1^{n-2}}{\varUpsilon _5^2}+\frac{24 \lambda  r^4 \psi _2^2 \psi _3^3 \varUpsilon _{15}^3}{\left(\varUpsilon _1^{3-n}+r^2 \psi _2 \psi _3 \varUpsilon _1\right){}^3},\\
\varUpsilon _{22}&&=\frac{r^2 \psi _3^2 \psi _2 \varUpsilon _{15}^2 \left(\frac{2 \lambda  \left((11 n+24) r^4 \psi _2^2+(48-11 n) r^2 \psi _2+24\right)}{\varUpsilon _1^2}+24\right)}{\left(\varUpsilon _1^{3-n}+r^2 \psi _2 \psi _3 \varUpsilon _1\right){}^2}-\frac{4 n \left(-\varUpsilon _1\right) \left(8 \lambda +\frac{2 \lambda  r^2 \psi _3 \psi _2 \varUpsilon _{16} \varUpsilon _1^{n-2}}{\varUpsilon _5^2}+4\right)}{\varUpsilon _1^2},\\
\varUpsilon _{23}&&=4 (\lambda +1) n \varUpsilon _1 \left(r^2 \psi _2-3\right)+\lambda  \left(24 n r^4 \psi _2^2-24 n r^2 \psi _2 \varUpsilon _1+3 n \varUpsilon _1^2\right.\\&&\left.+\frac{2 \psi _3 \varUpsilon _1^{n+1} \left(r^2 \psi _2 \psi _3^2 \left(\varUpsilon _1 \varUpsilon _8\right) \varUpsilon _1^{2 n}-\psi _3 \varUpsilon _7 \varUpsilon _1^{n+2}+(n-2) \varUpsilon _6 \varUpsilon _1^4\right)}{\varUpsilon _5^3}\right),
\end{eqnarray*}
\begin{eqnarray*}
\varUpsilon _{24}&&=\frac{2 r^2 \psi _2 \left(-\frac{7 \lambda  \psi _3 \varUpsilon _{16} \varUpsilon _1^{n+1}}{\varUpsilon _5^2}+4 n \left(r^4 \psi _2^2-1\right)+6 \lambda  n r^2 \psi _2 \left(r^2 \psi _2-3\right)\right)}{\varUpsilon _1^3}-10 \lambda,\\
\varUpsilon _{25}&&=\beta  \left(4 r^2 \psi _2 \left(-\frac{(\lambda -2) \psi _3 \varUpsilon _{16} \varUpsilon _1^{n-2}}{\varUpsilon _5^2}-\frac{2 (2 \lambda -1) n \left(-\varUpsilon _1\right)}{\varUpsilon _1^2}+\frac{(7 \lambda -3) r^2 \psi _2 \psi _3^2 \varUpsilon _{15}^2}{\left(\varUpsilon _1^{3-n}+r^2 \psi _2 \psi _3 \varUpsilon _1\right){}^2}\right.\right.\\&&\left.\left.+\frac{4 (\lambda +3) \psi _3 \varUpsilon _{15}}{\varUpsilon _1^{3-n}+r^2 \psi _2 \psi _3 \varUpsilon _1}\right)-8 \lambda \right)+2 \lambda  r^2 \varUpsilon _{14},\\
\varUpsilon _{26}&&=\frac{8 (\lambda -3) r^2 \psi _2 \psi _3^2 \varUpsilon _{15}^2}{\left(\varUpsilon _1^{3-n}+r^2 \psi _2 \psi _3 \varUpsilon _1\right){}^2}+\frac{4 (\lambda -2) n r^2 \psi _2 \left(r^2 \psi _2-3\right)}{\varUpsilon _1^3}-\frac{12 \lambda  r^4 \psi _3^3 \psi _2^2 \varUpsilon _{15}^3}{\left(\varUpsilon _1^{3-n}+r^2 \psi _2 \psi _3 \varUpsilon _1\right){}^3},\\
\varUpsilon _{27}&&=\frac{4 \lambda  r^2 \psi _3 \psi _2 \varUpsilon _1^{n-3} \left(r^2 \psi _2 \psi _3^2 \left(\varUpsilon _1 \varUpsilon _8\right) \varUpsilon _1^{2 n}-\psi _3 \varUpsilon _7 \varUpsilon _1^{n+2}+(n-2) \varUpsilon _6 \varUpsilon _1^4\right)}{\varUpsilon _5^3},\\
\varUpsilon _{28}&&=68 \lambda +\frac{2 r^2 \psi _2 \left(\frac{7 \lambda  \psi _3 \varUpsilon _{16} \varUpsilon _1^n}{\varUpsilon _5^2}+(13 \lambda -4) n \left(-\varUpsilon _1\right)\right)}{\varUpsilon _1^2}+32,
\end{eqnarray*}
\begin{eqnarray*}
\varUpsilon _{29}&&=\frac{4 \beta  (\lambda +1) n^4 r^6 \psi _2^3}{\varUpsilon _1^4}-\frac{44 \beta  (3 \lambda +2) n r^4 \psi _3^2 \psi _2^2 \left(-\varUpsilon _1\right) \varUpsilon _{15}^2 \varUpsilon _1^{2 n-4}}{\left(r^2 \psi _2 \left(\psi _3 \varUpsilon _1^n+2\right)+r^4 \psi _2^2+1\right){}^2}-\varUpsilon _2\\&&-\frac{2 (\lambda +1) n r^2 \left(-\varUpsilon _1\right) \left(r^2 \psi _2 \left(\psi _3 \varUpsilon _1^n+2\right)+r^4 \psi _2^2+1\right)}{\varUpsilon _1^4},\\
\varUpsilon _{30}&&=4 \beta  \psi _3 \varUpsilon _1^{n-2} \left(14 \lambda +(\lambda +1) r^2 \psi _2 \psi _3 \varUpsilon _1^{n-2}+8\right)-\frac{4 \beta  n^3 r^4 \psi _2^2 \left(\frac{2 (2 \lambda +3) r^2 \psi _3 \psi _2 \varUpsilon _{15}}{\varUpsilon _1^{3-n}+r^2 \psi _2 \psi _3 \varUpsilon _1}-2\right)}{\varUpsilon _1^3},\\
\varUpsilon _{31}&&=(7 \lambda +5) n^2 r^2 \psi _2 \left(-\varUpsilon _1\right){}^2+\frac{2 \lambda  \psi _3 \varUpsilon _{16} \varUpsilon _1^{n+2}}{\varUpsilon _5^2}+2 n \left(-\varUpsilon _1\right) \varUpsilon _1^2 \left(4 \lambda +\frac{2 (3 \lambda +2) r^2 \psi _3 \psi _2 \varUpsilon _{16} \varUpsilon _1^{n-2}}{\varUpsilon _5^2}+2\right)+2 r^2 \psi _2\\&&\times \left(2 (4 \lambda +3) n \varUpsilon _1 \left(r^2 \psi _2-3\right)-(3 \lambda +2) \left(24 n r^4 \psi _2^2-24 n r^2 \psi _2 \varUpsilon _1+3 n \varUpsilon _1^2\right.\right.\\&&\left.\left.+\frac{2 \psi _3 \varUpsilon _1^{n+1} \left(r^2 \psi _2 \psi _3^2 \left(\varUpsilon _1 \varUpsilon _8\right) \varUpsilon _1^{2 n}-\psi _3 \varUpsilon _7 \varUpsilon _1^{n+2}+(n-2) \varUpsilon _6 \varUpsilon _1^4\right)}{\varUpsilon _5^3}\right)\right),\\
\varUpsilon _{32}&&=4 \beta  (9 \lambda +7)+\frac{4 \beta  r^2 \psi _2 \left(2 (10 \lambda +7) n \left(-\varUpsilon _1\right) \varUpsilon _1+(3 \lambda +2) \left(\frac{7 \psi _3 \varUpsilon _{16} \varUpsilon _1^{n+1}}{\varUpsilon _5^2}-6 n r^2 \psi _2 \left(r^2 \psi _2-3\right)\right)\right)}{\varUpsilon _1^3}-\varUpsilon _{14}\\&&\times \left(12 \beta  (\lambda +1)+(2 \lambda +1) r^2\right),\\
\varUpsilon _{33}&&=-8 \beta  (3 \lambda +2)+4 \beta  r^2 \psi _2 \left(-\frac{(7 \lambda +6) \psi _3 \varUpsilon _{16} \varUpsilon _1^{n-2}}{\varUpsilon _5^2}-\frac{2 (2 \lambda +3) n \left(-\varUpsilon _1\right)}{\varUpsilon _1^2}+\frac{(21 \lambda +17) r^2 \psi _2 \psi _3^2 \varUpsilon _{15}^2}{\left(\varUpsilon _1^{3-n}+r^2 \psi _2 \psi _3 \varUpsilon _1\right){}^2}\right.\\&&-\left.\frac{4 (3 \lambda +4) \psi _3 \varUpsilon _{15}}{\varUpsilon _1^{3-n}+r^2 \psi _2 \psi _3 \varUpsilon _1}\right)+2 (\lambda +1) r^2 \varUpsilon _{14},\;\;\;\varUpsilon _{34}=\frac{2 (\lambda +1) r^4 \psi _3 \psi _2 \varUpsilon _{15}}{\varUpsilon _1^{3-n}+r^2 \psi _2 \psi _3 \varUpsilon _1}-2 \left(12 \beta  (\lambda +1)+r^2\right),
\end{eqnarray*}
\begin{eqnarray*}
\varUpsilon _{35}&&=4 (3 \lambda +2) r^2 \psi _3 \psi _2 \varUpsilon _1^{n-3} \left(r^2 \psi _2 \psi _3^2 \left(\varUpsilon _1 \varUpsilon _8\right) \varUpsilon _1^{2 n}-\psi _3 \varUpsilon _7 \varUpsilon _1^{n+2}+(n-2) \varUpsilon _6 \varUpsilon _1^4\right),\\
\varUpsilon _{36}&&=\psi _3 \varUpsilon _{15} \left(\frac{14 r^2 \psi _2 \left((5 \lambda +4) (-n) \left(-\varUpsilon _1\right)-\frac{(3 \lambda +2) \psi _3 \varUpsilon _{16} \varUpsilon _1^n}{\varUpsilon _5^2}\right)}{\varUpsilon _1^2}-4 (11 \lambda +6)\right).
\end{eqnarray*}

\section*{Appendix (\textbf{II})}

\begin{eqnarray*}
\lambda _1&&=-8 \psi _3 n \psi _2^8 r^{16} \varUpsilon _1^n-56 \psi _3 n \psi _2^7 r^{14} \varUpsilon _1^n-168 \psi _3 n \psi _2^6 r^{12} \varUpsilon _1^n-280 \psi _3 n \psi _2^5 r^{10} \varUpsilon _1^n-280 \psi _3 n \psi _2^4 r^8 \varUpsilon _1^n-168 \psi _3 n \psi _2^3 r^6 \varUpsilon _1^n\\&&-56 \psi _3 n \psi _2^2 r^4 \varUpsilon _1^n-8 \psi _3 n \psi _2 r^2 \varUpsilon _1^n+r^{16} \psi _2^8 \psi _3 \varUpsilon _1^n+8 r^{14} \psi _2^7 \psi _3 \varUpsilon _1^n+28 r^{12} \psi _2^6 \psi _3 \varUpsilon _1^n+56 r^{10} \psi _2^5 \psi _3 \varUpsilon _1^n+70 r^8 \psi _2^4 \psi _3 \varUpsilon _1^n\\&&+56 r^6 \psi _2^3 \psi _3 \varUpsilon _1^n+28 r^4 \psi _2^2 \psi _3 \varUpsilon _1^n+8 r^2 \psi _2 \psi _3 \varUpsilon _1^n+\psi _3 \varUpsilon _1^n-12 \psi _3^2 n \psi _2^7 r^{14} \varUpsilon _1^{2 n}-60 \psi _3^2 n \psi _2^6 r^{12} \varUpsilon _1^{2 n}-120 \psi _3^2 n \psi _2^5 r^{10} \varUpsilon _1^{2 n}\\&&-120 \psi _3^2 n \psi _2^4 r^8 \varUpsilon _1^{2 n}-60 \psi _3^2 n \psi _2^3 r^6 \varUpsilon _1^{2 n}-12 \psi _3^2 n \psi _2^2 r^4 \varUpsilon _1^{2 n}+4 r^{14} \psi _2^7 \psi _3^2 \varUpsilon _1^{2 n}+24 r^{12} \psi _2^6 \psi _3^2 \varUpsilon _1^{2 n}+60 r^{10} \psi _2^5 \psi _3^2 \varUpsilon _1^{2 n}\\&&+80 r^8 \psi _2^4 \psi _3^2 \varUpsilon _1^{2 n}+60 r^6 \psi _2^3 \psi _3^2 \varUpsilon _1^{2 n}+24 r^4 \psi _2^2 \psi _3^2 \varUpsilon _1^{2 n}+4 r^2 \psi _2 \psi _3^2 \varUpsilon _1^{2 n}-8 \psi _3^3 n \psi _2^6 r^{12} \varUpsilon _1^{3 n}-24 \psi _3^3 n \psi _2^5 r^{10} \varUpsilon _1^{3 n}\\&&-24 \psi _3^3 n \psi _2^4 r^8 \varUpsilon _1^{3 n}-8 \psi _3^3 n \psi _2^3 r^6 \varUpsilon _1^{3 n}+6 r^{12} \psi _2^6 \psi _3^3 \varUpsilon _1^{3 n}+24 r^{10} \psi _2^5 \psi _3^3 \varUpsilon _1^{3 n}+36 r^8 \psi _2^4 \psi _3^3 \varUpsilon _1^{3 n}+24 r^6 \psi _2^3 \psi _3^3 \varUpsilon _1^{3 n}\\&&+6 r^4 \psi _2^2 \psi _3^3 \varUpsilon _1^{3 n}+4 r^6 \psi _2^3 \psi _3^4 \varUpsilon _1^{4 n}-2 n \psi _2^9 r^{18}-18 n \psi _2^8 r^{16}-72 n \psi _2^7 r^{14}-168 n \psi _2^6 r^{12}-252 n \psi _2^5 r^{10}-252 n \psi _2^4 r^8\\&&-168 n \psi _2^3 r^6-72 n \psi _2^2 r^4-18 n \psi _2 r^2-2 n,
\end{eqnarray*}
\begin{eqnarray*}
\lambda _2&&=2 \beta  n^4 r^{16} \psi _2^9+12 \beta  n^4 r^{14} \psi _2^8+30 \beta  n^4 r^{12} \psi _2^7+40 \beta  n^4 r^{10} \psi _2^6+30 \beta  n^4 r^8 \psi _2^5+12 \beta  n^4 r^6 \psi _2^4+2 \beta  n^4 r^4 \psi _2^3+4 \beta  n^3 r^{16} \psi _2^9\\&&+20 \beta  n^3 r^{14} \psi _2^8+36 \beta  n^3 r^{12} \psi _2^7+20 \beta  n^3 r^{10} \psi _2^6-20 \beta  n^3 r^8 \psi _2^5-36 \beta  n^3 r^6 \psi _2^4-20 \beta  n^3 r^4 \psi _2^3+76 \beta  n^3 r^4 \psi _2^3 \psi _3 \varUpsilon _1^n\\&&-4 \beta  n^3 r^2 \psi _2^2-22 \beta  n^2 \psi _2+18 \beta  n^2 r^{16} \psi _2^9+128 \beta  n^2 r^{14} \psi _2^8+368 \beta  n^2 r^{12} \psi _2^7+528 \beta  n^2 r^{10} \psi _2^6+340 \beta  n^2 r^8 \psi _2^5-32 \beta  n^2 r^6\\&&\times \psi _2^4-192 \beta  n^2 r^4 \psi _2^3+660 \beta  n^2 r^4 \psi _2^3 \psi _3 \varUpsilon _1^n-112 \beta  n^2 r^2 \psi _2^2+130 \beta  n^2 r^2 \psi _2^2 \psi _3 \varUpsilon _1^n+80 \beta  n \psi _2-160 \beta  \psi _3 \psi _2 \varUpsilon _1^n+144 \beta  n \psi _2 \\&&\times\psi _3 \varUpsilon _1^n+16 \beta  n r^{16} \psi _2^9+192 \beta  n r^{14} \psi _2^8+896 \beta  n r^{12} \psi _2^7+2240 \beta  n r^{10} \psi _2^6-2 n r^{10} \psi _3^4 \psi _2^5 \varUpsilon _1^{4 n}+4 r^{10} \psi _2^5 \psi _3^4 \varUpsilon _1^{4 n}+3360 \\&&\times\beta  n r^8 \psi _2^5-2 n r^8 \psi _3^4 \psi _2^4 \varUpsilon _1^{4 n}+8 r^8 \psi _2^4 \psi _3^4 \varUpsilon _1^{4 n}+r^8 \psi _2^4 \psi _3^5 \varUpsilon _1^{5 n}+3136 \beta  n r^6 \psi _2^4-2720 \beta  r^6 \psi _3 \psi _2^4 \varUpsilon _1^n+1792 \beta  n r^4 \psi _2^3\\&&-2208 \beta  r^4 \psi _3 \psi _2^3 \varUpsilon _1^n+1840 \beta  n r^4 \psi _2^3 \psi _3 \varUpsilon _1^n+576 \beta  n r^2 \psi _2^2-928 \beta  r^2 \psi _3 \psi _2^2 \varUpsilon _1^n+800 \beta  n r^2 \psi _2^2 \psi _3 \varUpsilon _1^n
\end{eqnarray*}
\begin{eqnarray*}
\lambda _3&&=-480 \psi _3 \psi _2^6 r^{10} \beta  \varUpsilon _1^n-1760 \psi _3 \psi _2^5 r^8 \beta  \varUpsilon _1^n+18 n^4 r^{14} \beta  \psi _2^8 \psi _3 \varUpsilon _1^n+36 n^3 r^{14} \beta  \psi _2^8 \psi _3 \varUpsilon _1^n+18 n^2 r^{14} \beta  \psi _2^8 \psi _3 \varUpsilon _1^n+32 r^{14} \beta \\&&\times \psi _2^8 \psi _3 \varUpsilon _1^n+72 n^4 r^{12} \beta  \psi _2^7 \psi _3 \varUpsilon _1^n+220 n^3 r^{12} \beta  \psi _2^7 \psi _3 \varUpsilon _1^n+212 n^2 r^{12} \beta  \psi _2^7 \psi _3 \varUpsilon _1^n+80 n r^{12} \beta  \psi _2^7 \psi _3 \varUpsilon _1^n+32 r^{12} \beta  \psi _2^7 \psi _3 \varUpsilon _1^n\\&&+108 n^4 r^{10} \beta  \psi _2^6 \psi _3 \varUpsilon _1^n+520 n^3 r^{10} \beta  \psi _2^6 \psi _3 \varUpsilon _1^n+798 n^2 r^{10} \beta  \psi _2^6 \psi _3 \varUpsilon _1^n+544 n r^{10} \beta  \psi _2^6 \psi _3 \varUpsilon _1^n+72 n^4 r^8 \beta  \psi _2^5 \psi _3 \varUpsilon _1^n+600\\&&\times n^3 r^8 \beta  \psi _2^5 \psi _3 \varUpsilon _1^n+1432 n^2 r^8 \beta  \psi _2^5 \psi _3 \varUpsilon _1^n+1520 n r^8 \beta  \psi _2^5 \psi _3 \varUpsilon _1^n+18 n^4 r^6 \beta  \psi _2^4 \psi _3 \varUpsilon _1^n+340 n^3 r^6 \beta  \psi _2^4 \psi _3 \varUpsilon _1^n+1358 n^2 r^6 \beta  \\&&\times\psi _2^4 \psi _3 \varUpsilon _1^n+2240 n r^6 \beta  \psi _2^4 \psi _3 \varUpsilon _1^n+16 n^4 r^8 \beta  \psi _2^5 \psi _3^2 \varUpsilon _1^{2 n}+192 n^3 r^8 \beta  \psi _2^5 \psi _3^2 \varUpsilon _1^{2 n}+440 n^2 r^8 \beta  \psi _2^5 \psi _3^2 \varUpsilon _1^{2 n}+240 n r^8 \beta  \psi _2^5\\&&\times \psi _3^2 \varUpsilon _1^{2 n}+80 n^3 r^6 \beta  \psi _2^4 \psi _3^2 \varUpsilon _1^{2 n}+424 n^2 r^6 \beta  \psi _2^4 \psi _3^2 \varUpsilon _1^{2 n}+464 n r^6 \beta  \psi _2^4 \psi _3^2 \varUpsilon _1^{2 n}+144 n^2 r^4 \beta  \psi _2^3 \psi _3^2 \varUpsilon _1^{2 n}+392 n r^4 \beta  \psi _2^3 \psi _3^2 \varUpsilon _1^{2 n}\\&&+120 n r^2 \beta  \psi _2^2 \psi _3^2 \varUpsilon _1^{2 n}-284 \psi _3^2 \psi _2^6 r^{10} \beta  \varUpsilon _1^{2 n}-1110 \psi _3^2 \psi _2^5 r^8 \beta  \varUpsilon _1^{2 n}-1800 \psi _3^2 \psi _2^4 r^6 \beta  \varUpsilon _1^{2 n}-1430 \psi _3^2 \psi _2^3 r^4 \beta  \varUpsilon _1^{2 n}\\&&-540 \psi _3^2 \psi _2^2 r^2 \beta  \varUpsilon _1^{2 n}-74 \psi _3^2 \psi _2 \beta  \varUpsilon _1^{2 n}
\end{eqnarray*}
\begin{eqnarray*}
\lambda _4&&=16 \beta  n^4 r^{12} \psi _2^7 \psi _3^2 \varUpsilon _1^{2 n}+32 \beta  n^4 r^{10} \psi _2^6 \psi _3^2 \varUpsilon _1^{2 n}+32 \beta  n^3 r^{12} \psi _2^7 \psi _3^2 \varUpsilon _1^{2 n}+144 \beta  n^3 r^{10} \psi _2^6 \psi _3^2 \varUpsilon _1^{2 n}+24 \beta  n^2 r^{12} \psi _2^7 \psi _3^2 \varUpsilon _1^{2 n}+184 \\&&\times\beta  n^2 r^{10} \psi _2^6 \psi _3^2 \varUpsilon _1^{2 n}+24 \beta  n^2 r^{10} \psi _2^6 \psi _3^3 \varUpsilon _1^{3 n}+16 \beta  n^2 r^8 \psi _2^5 \psi _3^3 \varUpsilon _1^{3 n}-8 \beta  n^2 r^6 \psi _3^3 \psi _2^4 \varUpsilon _1^{3 n}+8 \beta  n r^{12} \psi _2^7 \psi _3^2 \varUpsilon _1^{2 n}-10 \beta  r^{12}\\&&\times \psi _3^2 \psi _2^7 \varUpsilon _1^{2 n}+56 \beta  n r^{10} \psi _2^6 \psi _3^2 \varUpsilon _1^{2 n}+24 \beta  n r^{10} \psi _2^6 \psi _3^3 \varUpsilon _1^{3 n}-62 \beta  r^{10} \psi _3^3 \psi _2^6 \varUpsilon _1^{3 n}+56 \beta  n r^8 \psi _2^5 \psi _3^3 \varUpsilon _1^{3 n}-248 \beta  r^8 \psi _3^3 \psi _2^5 \varUpsilon _1^{3 n}\\&&-22 \beta  r^8 \psi _3^4 \psi _2^5 \varUpsilon _1^{4 n}+8 \beta  n r^6 \psi _2^4 \psi _3^3 \varUpsilon _1^{3 n}-404 \beta  r^6 \psi _3^3 \psi _2^4 \varUpsilon _1^{3 n}-44 \beta  r^6 \psi _3^4 \psi _2^4 \varUpsilon _1^{4 n}-2 \beta  r^6 \psi _3^5 \psi _2^4 \varUpsilon _1^{5 n}-312 \beta  r^4 \psi _3^3 \psi _2^3 \varUpsilon _1^{3 n}\\&&-24 \beta  n r^4 \psi _3^3 \psi _2^3 \varUpsilon _1^{3 n}-22 \beta  r^4 \psi _3^4 \psi _2^3 \varUpsilon _1^{4 n}-94 \beta  r^2 \psi _3^3 \psi _2^2 \varUpsilon _1^{3 n}
\end{eqnarray*}
\begin{eqnarray*}
\lambda _5&&=-3 \psi _3 n^2 \psi _2^8 r^{16} \varUpsilon _1^n-2 \psi _3 \psi _2^8 r^{16} \varUpsilon _1^n-18 \psi _3 n^2 \psi _2^7 r^{14} \varUpsilon _1^n-16 \psi _3 \psi _2^7 r^{14} \varUpsilon _1^n-45 \psi _3 n^2 \psi _2^6 r^{12} \varUpsilon _1^n-56 \psi _3 \psi _2^6 r^{12} \varUpsilon _1^n\\&&-60 \psi _3 n^2 \psi _2^5 r^{10} \varUpsilon _1^n-112 \psi _3 \psi _2^5 r^{10} \varUpsilon _1^n-45 \psi _3 n^2 \psi _2^4 r^8 \varUpsilon _1^n-140 \psi _3 \psi _2^4 r^8 \varUpsilon _1^n-18 \psi _3 n^2 \psi _2^3 r^6 \varUpsilon _1^n-112 \psi _3 \psi _2^3 r^6 \varUpsilon _1^n\\&&-3 \psi _3 n^2 \psi _2^2 r^4 \varUpsilon _1^n-56 \psi _3 \psi _2^2 r^4 \varUpsilon _1^n-16 \psi _3 \psi _2 r^2 \varUpsilon _1^n+11 n r^{16} \psi _2^8 \psi _3 \varUpsilon _1^n+71 n r^{14} \psi _2^7 \psi _3 \varUpsilon _1^n+195 n r^{12} \psi _2^6 \psi _3 \varUpsilon _1^n+295 \\&&\times n r^{10} \psi _2^5 \psi _3 \varUpsilon _1^n+265 n r^8 \psi _2^4 \psi _3 \varUpsilon _1^n+141 n r^6 \psi _2^3 \psi _3 \varUpsilon _1^n+41 n r^4 \psi _2^2 \psi _3 \varUpsilon _1^n+5 n r^2 \psi _2 \psi _3 \varUpsilon _1^n-2 \psi _3 \varUpsilon _1^n-120 \psi _3^2 \psi _2^5 r^{10} \varUpsilon _1^{2 n}\\&&-12 \psi _3^2 n^2 \psi _2^4 r^8 \varUpsilon _1^{2 n}-160 \psi _3^2 \psi _2^4 r^8 \varUpsilon _1^{2 n}-3 \psi _3^2 n^2 \psi _2^3 r^6 \varUpsilon _1^{2 n}-120 \psi _3^2 \psi _2^3 r^6 \varUpsilon _1^{2 n}-48 \psi _3^2 \psi _2^2 r^4 \varUpsilon _1^{2 n}-8 \psi _3^2 \psi _2 r^2 \varUpsilon _1^{2 n}+114\\&&\times  n r^8 \psi _2^4 \psi _3^2 \varUpsilon _1^{2 n}+51 n r^6 \psi _2^3 \psi _3^2 \varUpsilon _1^{2 n}+9 n r^4 \psi _2^2 \psi _3^2 \varUpsilon _1^{2 n}-n^2 \psi _2^9 r^{18}-8 n^2 \psi _2^8 r^{16}-28 n^2 \psi _2^7 r^{14}-56 n^2 \psi _2^6 r^{12}-70 n^2 \psi _2^5\\&&\times r^{10}+3 n r^{18} \psi _2^9-56 n^2 \psi _2^4 r^8+25 n r^{16} \psi _2^8+92 n r^{14} \psi _2^7-28 n^2 \psi _2^3 r^6+196 n r^{12} \psi _2^6+266 n r^{10} \psi _2^5-8 n^2 \psi _2^2 r^4\\&&+238 n r^8 \psi _2^4+140 n r^6 \psi _2^3-n^2 \psi _2 r^2+52 n r^4 \psi _2^2+n+11 n r^2 \psi _2
\end{eqnarray*}
\begin{eqnarray*}
\lambda _6&&=-180 \psi _3 n \psi _2 \beta  \varUpsilon _1^n+200 \beta  \psi _2 \psi _3 \varUpsilon _1^n-3 \psi _3^2 n^2 \psi _2^7 r^{14} \varUpsilon _1^{2 n}-8 \psi _3^2 \psi _2^7 r^{14} \varUpsilon _1^{2 n}-12 \psi _3^2 n^2 \psi _2^6 r^{12} \varUpsilon _1^{2 n}-48 \psi _3^2 \psi _2^6 r^{12} \varUpsilon _1^{2 n}\\&&-18 \psi _3^2 n^2 \psi _2^5 r^{10} \varUpsilon _1^{2 n}+15 n r^{14} \psi _2^7 \psi _3^2 \varUpsilon _1^{2 n}+69 n r^{12} \psi _2^6 \psi _3^2 \varUpsilon _1^{2 n}+126 n r^{10} \psi _2^5 \psi _3^2 \varUpsilon _1^{2 n}-12 \psi _3^3 \psi _2^6 r^{12} \varUpsilon _1^{3 n}-\psi _3^3 n^2 \psi _2^6 r^{12} \varUpsilon _1^{3 n}\\&&-48 \psi _3^3 \psi _2^5 r^{10} \varUpsilon _1^{3 n}-2 \psi _3^3 n^2 \psi _2^5 r^{10} \varUpsilon _1^{3 n}-72 \psi _3^3 \psi _2^4 r^8 \varUpsilon _1^{3 n}-\psi _3^3 n^2 \psi _2^4 r^8 \varUpsilon _1^{3 n}-48 \psi _3^3 \psi _2^3 r^6 \varUpsilon _1^{3 n}-12 \psi _3^3 \psi _2^2 r^4 \varUpsilon _1^{3 n}+9 n r^{12}\\&&\times \psi _2^6 \psi _3^3 \varUpsilon _1^{3 n}+25 n r^{10} \psi _2^5 \psi _3^3 \varUpsilon _1^{3 n}+23 n r^8 \psi _2^4 \psi _3^3 \varUpsilon _1^{3 n}+7 n r^6 \psi _2^3 \psi _3^3 \varUpsilon _1^{3 n}-8 \psi _3^4 \psi _2^5 r^{10} \varUpsilon _1^{4 n}-16 \psi _3^4 \psi _2^4 r^8 \varUpsilon _1^{4 n}-8 \psi _3^4 \psi _2^3 r^6 \varUpsilon _1^{4 n}\\&&+2 n r^{10} \psi _2^5 \psi _3^4 \varUpsilon _1^{4 n}+2 n r^8 \psi _2^4 \psi _3^4 \varUpsilon _1^{4 n}-2 \psi _3^5 \psi _2^4 r^8 \varUpsilon _1^{5 n}+84 n^2 r^{10} \beta  \psi _2^6+560 n^2 r^8 \beta  \psi _2^5+700 n^2 r^6 \beta  \psi _2^4+420 n^2 r^4 \beta  \psi _2^3\\&&+124 n^2 r^2 \beta  \psi _2^2-6 n^4 \psi _2^9 r^{16} \beta -30 n^2 \psi _2^9 r^{16} \beta -36 n \psi _2^9 r^{16} \beta -36 n^4 \psi _2^8 r^{14} \beta -16 n^3 \psi _2^8 r^{14} \beta -140 n^2 \psi _2^8 r^{14} \beta -448 n\\&&\times \psi _2^8 r^{14} \beta -90 n^4 \psi _2^7 r^{12} \beta -96 n^3 \psi _2^7 r^{12} \beta -196 n^2 \psi _2^7 r^{12} \beta -2032 n \psi _2^7 r^{12} \beta -120 n^4 \psi _2^6 r^{10} \beta -240 n^3 \psi _2^6 r^{10} \beta -4800 n\\&&\times \psi _2^6 r^{10} \beta -90 n^4 \psi _2^5 r^8 \beta -320 n^3 \psi _2^5 r^8 \beta -6680 n \psi _2^5 r^8 \beta -36 n^4 \psi _2^4 r^6 \beta -240 n^3 \psi _2^4 r^6 \beta -5696 n \psi _2^4 r^6 \beta -6 n^4 \psi _2^3 r^4 \beta\\&& -96 n^3 \psi _2^3 r^4 \beta -2928 n \psi _2^3 r^4 \beta -16 n^3 \psi _2^2 r^2 \beta -832 n \psi _2^2 r^2 \beta -100 n \psi _2 \beta +14 n^2 \beta  \psi _2
\end{eqnarray*}
\begin{eqnarray*}
\lambda _7&&=-104 \psi _3 n^3 \psi _2^8 r^{14} \beta  \varUpsilon _1^n-90 \psi _3 n^2 \psi _2^8 r^{14} \beta  \varUpsilon _1^n-88 \psi _3 \psi _2^8 r^{14} \beta  \varUpsilon _1^n-376 \psi _3 n^3 \psi _2^7 r^{12} \beta  \varUpsilon _1^n-840 \psi _3 n^2 \psi _2^7 r^{12} \beta  \varUpsilon _1^n\\&&-12 \psi _3 n \psi _2^7 r^{12} \beta  \varUpsilon _1^n-464 \psi _3 n^3 \psi _2^6 r^{10} \beta  \varUpsilon _1^n-2426 \psi _3 n^2 \psi _2^6 r^{10} \beta  \varUpsilon _1^n-1848 \psi _3 n \psi _2^6 r^{10} \beta  \varUpsilon _1^n-176 \psi _3 n^3 \psi _2^5 r^8 \beta  \varUpsilon _1^n-3104\\&&\times \psi _3  n^2 \psi _2^5 r^8 \beta  \varUpsilon _1^n-5692 \psi _3 n \psi _2^5 r^8 \beta  \varUpsilon _1^n-1806 \psi _3 n^2 \psi _2^4 r^6 \beta  \varUpsilon _1^n-7608 \psi _3 n \psi _2^4 r^6 \beta  \varUpsilon _1^n-344 \psi _3 n^2 \psi _2^3 r^4 \beta  \varUpsilon _1^n-5124 \psi _3 n\\&&\times \psi _2^3 r^4 \beta  \varUpsilon _1^n-1640 \psi _3 n \psi _2^2 r^2 \beta  \varUpsilon _1^n+2 n^4 r^{14} \beta  \psi _2^8 \psi _3 \varUpsilon _1^n+88 n r^{14} \beta  \psi _2^8 \psi _3 \varUpsilon _1^n+8 n^4 r^{12} \beta  \psi _2^7 \psi _3 \varUpsilon _1^n+56 r^{12} \beta  \psi _2^7 \psi _3 \varUpsilon _1^n\\&&+12 n^4 r^{10} \beta  \psi _2^6 \psi _3 \varUpsilon _1^n+1800 r^{10} \beta  \psi _2^6 \psi _3 \varUpsilon _1^n+8 n^4 r^8 \beta  \psi _2^5 \psi _3 \varUpsilon _1^n+5080 r^8 \beta  \psi _2^5 \psi _3 \varUpsilon _1^n+2 n^4 r^6 \beta  \psi _2^4 \psi _3 \varUpsilon _1^n+56 n^3 r^6 \beta  \psi _2^4 \psi _3\\&&\times \varUpsilon _1^n+6520 r^6 \beta  \psi _2^4 \psi _3 \varUpsilon _1^n+40 n^3 r^4 \beta  \psi _2^3 \psi _3 \varUpsilon _1^n+4392 r^4 \beta  \psi _2^3 \psi _3 \varUpsilon _1^n+34 n^2 r^2 \beta  \psi _2^2 \psi _3 \varUpsilon _1^n+1496 r^2 \beta  \psi _2^2 \psi _3 \varUpsilon _1^n+3454 r^4\\&&\times \beta  \psi _2^3 \psi _3^2 \varUpsilon _1^{2 n}+1220 r^2 \beta  \psi _2^2 \psi _3^2 \varUpsilon _1^{2 n}+106 \beta  \psi _2 \psi _3^2 \varUpsilon _1^{2 n}-468 \psi _3^2 n^2 \psi _2^3 r^4 \beta  \varUpsilon _1^{2 n}-1408 \psi _3^2 n \psi _2^3 r^4 \beta  \varUpsilon _1^{2 n}-520 \psi _3^2 n \psi _2^2 r^2 \beta  \varUpsilon _1^{2 n}
\end{eqnarray*}
\begin{eqnarray*}
\lambda _8&&=60 n^2 r^{12} \beta  \psi _2^7 \psi _3^2 \varUpsilon _1^{2 n}+114 r^{12} \beta  \psi _2^7 \psi _3^2 \varUpsilon _1^{2 n}+772 r^{10} \beta  \psi _2^6 \psi _3^2 \varUpsilon _1^{2 n}+2534 r^8 \beta  \psi _2^5 \psi _3^2 \varUpsilon _1^{2 n}+4216 r^6 \beta  \psi _2^4 \psi _3^2 \varUpsilon _1^{2 n}-48 \psi _3^2 n^4\\&&\times \psi _2^7 r^{12} \beta  \varUpsilon _1^{2 n}-48 \psi _3^2 n^3 \psi _2^7 r^{12} \beta  \varUpsilon _1^{2 n}-216 \psi _3^2 n \psi _2^7 r^{12} \beta  \varUpsilon _1^{2 n}-96 \psi _3^2 n^4 \psi _2^6 r^{10} \beta  \varUpsilon _1^{2 n}-320 \psi _3^2 n^3 \psi _2^6 r^{10} \beta  \varUpsilon _1^{2 n}-96 \psi _3^2 n^2\\&&\times \psi _2^6 r^{10} \beta  \varUpsilon _1^{2 n}-560 \psi _3^2 n \psi _2^6 r^{10} \beta  \varUpsilon _1^{2 n}-48 \psi _3^2 n^4 \psi _2^5 r^8 \beta  \varUpsilon _1^{2 n}-496 \psi _3^2 n^3 \psi _2^5 r^8 \beta  \varUpsilon _1^{2 n}-840 \psi _3^2 n^2 \psi _2^5 r^8 \beta  \varUpsilon _1^{2 n}-840 \psi _3^2 n \\&&\times\psi _2^5 r^8 \beta  \varUpsilon _1^{2 n}-224 \psi _3^2 n^3 \psi _2^4 r^6 \beta  \varUpsilon _1^{2 n}-1152 \psi _3^2 n^2 \psi _2^4 r^6 \beta  \varUpsilon _1^{2 n}-1384 \psi _3^2 n \psi _2^4 r^6 \beta  \varUpsilon _1^{2 n}+126 r^{10} \beta  \psi _2^6 \psi _3^3 \varUpsilon _1^{3 n}+528 r^8 \beta \\&&\times \psi _2^5 \psi _3^3 \varUpsilon _1^{3 n}+16 n^2 r^6 \beta  \psi _2^4 \psi _3^3 \varUpsilon _1^{3 n}+924 r^6 \beta  \psi _2^4 \psi _3^3 \varUpsilon _1^{3 n}+52 n r^4 \beta  \psi _2^3 \psi _3^3 \varUpsilon _1^{3 n}+800 r^4 \beta  \psi _2^3 \psi _3^3 \varUpsilon _1^{3 n}+278 r^2 \beta  \psi _2^2 \psi _3^3 \varUpsilon _1^{3 n}\\&&-48 \psi _3^3 n^2 \psi _2^6 r^{10} \beta  \varUpsilon _1^{3 n}-60 \psi _3^3 n \psi _2^6 r^{10} \beta  \varUpsilon _1^{3 n}-32 \psi _3^3 n^2 \psi _2^5 r^8 \beta  \varUpsilon _1^{3 n}-164 \psi _3^3 n \psi _2^5 r^8 \beta  \varUpsilon _1^{3 n}-52 \psi _3^3 n \psi _2^4 r^6 \beta  \varUpsilon _1^{3 n}+42 r^8 \beta \\&&\times \psi _2^5 \psi _3^4 \varUpsilon _1^{4 n}+108 r^6 \beta  \psi _2^4 \psi _3^4 \varUpsilon _1^{4 n}+66 r^4 \beta  \psi _2^3 \psi _3^4 \varUpsilon _1^{4 n}+6 r^6 \beta  \psi _2^4 \psi _3^5 \varUpsilon _1^{5 n}
\end{eqnarray*}


\section*{References}
\begin{thebibliography}{}
\bibitem{1} Baade and Zwicky., PNAS \textbf{112}, 1241 (2015).
\bibitem{2} K.R. Karmarkar, Proc. Indian Acad. Sci. A \textbf{27}, 56 (1948).
\bibitem{3} L. Schlai, Ann. Mat. \textbf{5}, 170 (1871).
\bibitem{4} J. Nash, Ann. of Math. \textbf{63}, 20 (1956).
\bibitem{5} S.K. Maurya et al., Eur. Phys. J. C \textbf{75}, 389 (2015).
\bibitem{6} S.K. Maurya et al., Eur. Phys. J. A \textbf{52} 191 (2016).
\bibitem{7} P. Bhar et al., Eur. Phys. J. A \textbf{52}, 312 (2016).
\bibitem{8} D. Deb, S.V. Ketov, S.K. Maurya, Mon. Not. R. Astron. Soc. \textbf{485}, 5652 (2019).
\bibitem{9} S.K. Maurya, et al., Phys. Rev. D \textbf{100}, (2019).
\bibitem{10} S.K. Maurya et al. , Ann. Physics \textbf{385},  532 (2017).
\bibitem{11} S.K. Maurya et al., Eur. Phys. J. C \textbf{76}, 266 (2016).
\bibitem{12} S.K. Maurya et al., Eur. Phys. J. C \textbf{77}, 1 (2016).
\bibitem{13} T. Harko et al., Phys. Rev. D \textbf{84}, 024020 (2011).
\bibitem{14} M. Jamil, D. Momeni and R. Myrzakulov., Eur. Phys. J. C. \textbf{72}, 1959 (2012).
\bibitem{15} M. Jamil, D. Momeni and R. Myrzakulov., Chin. Phys. Lett. \textbf{29}, 109801 (2012).
\bibitem{16} H. Shabani and M. Farhoudi.,  Chin. Phys. Lett. \textbf{88}, 044048 (2013).
\bibitem{17} H. Shabani and M. Farhoudi.,  Phys. Rev. D  \textbf{90}, 44031 (2014).
\bibitem{18} P.H.R.S. Moraes .,  Eur. Phys. J. C  \textbf{75}, 168 (2015).
\bibitem{19} A. Alhamzawi and R. Alhamzawi.,  Int. J. Mod. Phys. D \textbf{25}, 1650020 (2016).
\bibitem{20} P.H.R.S. Moraes, J.D. Arbanil and M. Malheiro., J. Cosmol. Astropart. Phys. \textbf{6}, 5 (2016).

\bibitem{21} Das et al., Eur. Phys. J. C \textbf{76}, 654 (2016).
\bibitem{22} Moraes, al., arXiv:1806.04123v4.
\bibitem{23} Z. Yousaf, M.Z. Bhatti and M. Ilyas ., Eur. Phys. J. C \textbf{78}, 307 (2018).
\bibitem{24} S.K.Mauryaa and F. T. Ortizb., Phys. Dark Universe \textbf{27}, 100442 (2020).
\bibitem{25} S. Waheed., Symmetry \textbf{12}, 962 (2020).
\bibitem{26} G. Mustafa et al., Eur. Phys. J. C \textbf{80}, 26 (2020).
\bibitem{27} M. Ruderman., Annu Rev Astron Astr, \textbf{10}, 427 (1972).
\bibitem{28} A. V. Astashenok, S. Capozziello and S. D. Odintsov., J. Cosmol. Astropart. Phys. \textbf{01}, 001 (2015).
\bibitem{29} A. V. Astashenok, S. Capozziello and S. D. Odintsov.,  Phys. Lett. B \textbf{742}, 160 (2015).
\bibitem{30} D. Momeni, P. H. R. S. Moraes and R. Myrzakulov., Astrophys. Space Sci \textbf{361}, 228 (2018).
\bibitem{31} D. Momeni, M. Raza and R. Myrzakulov.,  Mod. Phys. Lett. A \textbf{31}, 1650073 (2015).
\bibitem{32} S. Capozziello et al., Phys. Rev. D \textbf{93}, 023501 (2016).
\bibitem{33} A. V. Astashenok, S. Capozziello and S. D. Odintsov., Phys. Rev. D \textbf{89}, 103509 (2014).
\bibitem{Herrera}L. Herrera, Phys. Lett. A \textbf{165}, 206 (1992).
\bibitem{34} A. A. Starobinsky, Phys. Lett. B \textbf{91}, 99 (1980).
\bibitem{35} M. Jamil et al., Eur. Phys. J. C \textbf{72}, 1999 (2012).
\bibitem{36} H. Shabani and M. Farhoudi, Phys. Rev. D \textbf{88}, 044048 (2013).
\bibitem{37} C.P. Singh and P. Kumar, Eur. Phys. J. C \textbf{74}, 3070 (2014).
\bibitem{38} M. Sharif, Z. Yousaf, Astrophys. Space Sci. \textbf{354}, 471 (2014).
\bibitem{39} I. Noureen and M. Zubair, Astrophys. Space Sci. \textbf{356}, 103 (2015).
\bibitem{40} I. Noureen et al., Eur. Phys. J. C \textbf{75}, 323 (2015).
\bibitem{41} S. Chandrasekhar, Astrophys. J. \textbf{140}, 417 (1964).
\bibitem{42} H. Heintzmann and W. Hillebrandt, Astron. Astrophys.\textbf{ 38}, 51 (1975).
\bibitem{43} W. Hillebrandt and K. O. Steinmetz, Astron. Astrophys. \textbf{53}, 283 (1976).
\bibitem{44} D. Horvat, S. Ilijic and  A. Marunovic, Class. Quantum Grav. \textbf{28}, 025009 (2011).
\bibitem{45} D. D. Doneva and S. S. Yazadjiev, Phys. Rev. D \textbf{85}, 124023 (2012).
\bibitem{46} H. O. Silva, C. F. B. Macedo, E. Berti and L. C. B. Crispino, Class. Quant. Grav. \textbf{32}, 145008 (2015).
\bibitem{47} I. Bombaci, Astron. Astrophys. \textbf{305}, 871 (1996).

\bibitem{48} M. Zubair, G. Abbas, and I. Noureen, Astrophys. Space Sci. \textbf{361},8 (2016).
\bibitem{49} A. K. Yadav, M. Mondal, and F. Rahaman,  Pramana - J Phys \textbf{94}, 90 (2020).

\end{thebibliography}
\end{document}